\documentclass[twocolumn,aps,pra]{revtex4}
\usepackage{epsfig}
\usepackage[english]{babel}
\usepackage{latexsym}
\usepackage{graphics}
\usepackage{subfigure}
\usepackage{epsfig}
\usepackage{graphicx}
\usepackage{dcolumn}
\usepackage{amsmath}
\usepackage{hyperref}
\usepackage{amssymb}
\usepackage{xcolor}
\begin{document}
\title{Local most-probable routes and classic-quantum correspondence in strong-field two-dimensional tunneling ionization}
\author{C. Chen$^{1}$, X. X. Ji$^{1}$, W. Y. Li$^{2,\dag}$, X. Han$^{1}$,  and Y. J. Chen$^{1,*}$}

\date{\today}

\begin{abstract}
We study ionization of atoms in strong two-dimensional (2D) laser fields with various forms, numerically and analytically.
We focus on the local most-probable tunneling routes (some specific electron trajectories)
which are corresponding to the local maxima of photoelectron momentum distributions (PMDs).
By making classic-quantum correspondence, we obtain a condition for these routes characterized by the electron position at the tunnel exit.
With comparing the identified routes with the classical limit
and the partial-decoupling approximation where it is assumed that tunneling is dominated by the main component of the 2D field,
some semiclassical properties of 2D tunneling are addressed.
The local maxima of PMD related to the local most-probable routes can be used as one of the preferred observables in ultrafast measurements.
\end{abstract}

\affiliation{
1.College of Physics and Information Technology, Shaanxi Normal University,Xi'an710119, China\\
2.School of Mathematics and Science, Hebei GEO University, Shijiazhuang 050031, China}

\maketitle

\section{Introduction}
When an atom or a molecule is exposed to the strong laser field, the coulomb potential is bent by the electric field,
with forming a barrier of which the electron wave packet (EWP) can tunnel out.
The laser-induced tunneling ionization is the first step of a broad range of strong-field ultrafast processes,
such as above-threshold ionization (ATI) \cite{Agostini1979, Yang1993, Paulus1994, Lewenstein1995},
high-order harmonic generation (HHG) \cite{McPherson1987, Huillier1991, Corkum1993, Krausz2009},
and nonsequential double ionization (NSDI) \cite{Niikura2003, Zeidler2005, Becker2012}.
These processes can be well understood with strong-field approximations (SFA) \cite{Keldysh,Faisal,Reiss,Lewenstein1994,Milosevic2006}.
The use of saddle-point theory \cite{Lewenstein1994,Milosevic2006} in SFA also allows one to describe the ionized EWPs in form of the complex electron trajectory,
giving an intuitive semiclassical physical picture of these processes.
The ionized EWPs carry the phase and structural information of the bound electronic states as well as the instantaneous information of the electric field
at the time of tunneling.
Therefore, the photoelectron momentum distribution (PMD) of atoms or  molecules, associated with ionized EWPs,
can be used to probe the electronic structure \cite{Meckel2008, Murray2011, Gao2017},
the ionization dynamics \cite{Xie2012, Wang2017}, and the Coulomb effect \cite{Huismans2011, TMYan2010}.

Recently, there is an increasing interest in the study of strong-field ionization with two-dimensional (2D) laser fields
such as orthogonally-polarized two-color (OTC) laser fields and elliptically-polarized ones.
The 2D property of these fields opens up new perspectives for controlling and probing the electron motion on the atomic time scale,
and many new phenomena and effects associated with tunneling are revealed in relevant studies.
For instance, the OTC laser field has been used to control the electron recollision in atomic and molecular HHG \cite{Kitzler2005,Brugnera2011,chen2018},
image the electronic structure within atoms and molecules \cite{Kitzler2007,Shafir2009,Das2013},
resolve the time when the electron tunnels out of a barrier \cite{Shafir2,Lein2013},
control the interference between EWPs released at different times within one laser cycle \cite{Richter2015},
separate the quantum paths in the multiphoton ionization of atoms \cite{Xinhua2017},
probe the nonadiabatic effect and the effect of the sub-barrier phase on tunneling electrons \cite{Geng2015, Meng Han2017},
distinguish the coulomb-modified trajectory contributions to tunneling \cite{SGYu2016},
and identify the Coulomb-induced ionization time lag after tunneling \cite{chen2019}, etc..
The elliptical laser field has been used as an ``attosecond clock" to timing the electron motion during the tunneling process \cite{Eckle1,Eckle2} and
calibrate nonadiabatic tunneling effects \cite{Pfeiffer,Klaiber,Luo2019}, etc..

Because of the importance of 2D laser fields in ultrafast measurements and controls of electron motion, in this paper,
we focus on some basic characteristics of 2D tunneling such as the local most-probable tunneling routes \cite{Luo2019,chenchao2019,XieW2019},
corresponding to the regions with the local maximal amplitude in PMDs,
and explore their semiclassical properties and their transitions towards the classical limit,
which have not been sufficiently addressed before and
will provide insight into the complex dynamics of atoms or molecules in strong 2D laser fields.
One of these local routes, i.e., the most-probable route related to the region of the maximal amplitude in PMD,
has shown important applications in ultrafast measurements.
The use of the distribution of the most-probable route as the observable in ultrafast measurements, instead of using the total PMD,
allows one to explore some issues with high time resolution,
such as the attosecond-scale issues of tunneling delay time \cite{Eckle1} and ionization time shift \cite{XieW2019}.

Based on the numerical solution of the time-dependent Schr\"{o}dinger equation (TDSE) and SFA,
we study ATI of a model He atom in 2D laser fields with diverse forms.
We first explore a general condition for identifying the local most-probable routes associated with specific electron trajectories in 2D tunneling.
By analyzing classic-quantum correspondence at the tunneling exit, we obtain the relationship between the  exit position and the local most-probable route.
Then using the OTC laser field with zero time delay between these two colors as a case,
we compare the predictions of the local most-probable route with the classical predictions and the predictions of the partial-decoupling approximation,
where the main component of the 2D field with large laser amplitude is considered to dominate in tunneling.
We further extend our discussions to cases of 2D laser fields with other forms such as OTC laser fields with large time delay,
elliptical and circular ones to validate our conclusions.
With the help of the Coulomb-modified SFA model, effects of Coulomb potential on the local most-probable tunneling route are addressed.
Finally, with using circular laser fields, we discuss the potential applications of the local most-probable route in ultrafast measurements.

\section{Theory methods}
\subsection{Numerical methods}
In the length gauge and the dipole approximation, the Hamiltonian of the model He atom interacting with a strong laser field can be written as
\begin{eqnarray}
H(t)=-\frac{\nabla ^2}{2}+V(\mathbf{r})+\mathbf{r}\cdot\mathbf{E}(t),
\end{eqnarray}
(in atomic units of $\hbar=e=m_e=1$).
Here, the term $V(\mathbf{r})=-{Z}/{\sqrt{\xi+x^2+z^2}}$ represents the Coulomb potential of the model atom,
$\xi=0.5$ is the smoothing parameter which is used to avoid the Coulomb singularity,
and $Z$ is the effective nuclear charge which is adjusted such that the ionization potential $I_p$ of the model system reproduced here is $I_p=0.9$ a.u..

In our simulations, we assume that the atom is located in the origin and the orthogonally-polarized 2D laser field is located in the $xz$ plane
with its two polarization components along the $x$ axis and the $z$ axis, respectively.

For the first case of the OTC laser field explored in the paper, we assume that
the fundamental field is along the $x$ axis and the additional second-harmonic field is along the $z$ axis.
In this way, the electric field of OTC can be written as
\begin{eqnarray}
\mathbf{E}(t)=f(t)[E_1\sin(\omega_1t)\mathbf{e}_x+E_2\sin(\omega_2t+\phi)\mathbf{e}_z].
\end{eqnarray}
The symbol $\mathbf{e}_x(\mathbf{e}_z)$ denotes the unit vector along the $x(z)$ axis.
$E_1$  is the maximal laser amplitude relating to the peak intensity $I_1$ of the fundamental field
and $E_2=\varepsilon E_1$ is that of the second-harmonic field. $\varepsilon$ is the ratio of  $E_2$ to $E_1$.
$\omega_1$ and $\omega_2=2\omega_1$ are the laser frequencies of the fundamental and second-harmonic fields.
$\phi$ is the relative phase and $f(t)$ is the envelope function.
In our calculations, we work with a twenty-cycle laser pulse which is linearly ramped up for three optical cycles,
then kept at a constant intensity for fourteen optical cycles, and finally linearly ramped down for three optical cycles.

The TDSE is solved numerically using the spectral method \cite{Feit1982}.
We use a grid size of $L_x \times L_z=410\ \text{a.u.}\times410\ \text{a.u.}$ with a grid spacing of $\Delta x=\Delta z=0.4\ \text{a.u.}$.
The time step is $\Delta t=0.05\ \text{a.u.}$.
For each time interval of $1 \text{a.u.}$, the TDSE wave function $\psi(t)$ of $H(t)$ is multiplied by a $\cos^{1/2}$ mask function
to absorb the continuum wave packet at the boundary from which we obtain PMDs \cite{Gao2017}.

Unless mentioned otherwise, we use the laser parameters of $I_1=5\times 10^{14} \text{W/cm}^2$, $\lambda_1=800$ nm ($\omega_1=0.057$ a.u.),
$\varepsilon=0.5$  and $\phi=0$.

\subsection{Strong-field approximation}
In the frame of SFA, the transition amplitude of the photoelectron with the drift momentum $\mathbf{p}$ can be written as \cite{Lewenstein1994,Milosevic2006}
\begin{eqnarray}
M_\mathbf{p}=-i\int _{0}^{t_F}dt' {\mathbf{E}(t')}\cdot\langle {\mathbf{p}+\mathbf{A}(t')} |{\mathbf{r}}| 0 \rangle {e^{iS(\mathbf{p},t')}},
\end{eqnarray}
where
\begin{eqnarray}
S(\mathbf{p},t')=-\int _{t'}^{t_F}\{\frac{[\mathbf{p}+\mathbf{A}(t'')]^2}{2}+I_p\}dt''
\end{eqnarray}
is the semiclassical action, which corresponds to the propagation of an electron from the ionization time $t'$ to the final time $t_F$.
Here, $\mathbf{A}(t)=-\int^t\mathbf{E}(t')dt'$ is the vector potential of the laser field $\mathbf{E}(t)$. $I_p$ is the ionization potential.
$|0\rangle$ denotes the initial-state wave function of the atom.
The term $\langle {\mathbf{p}+\mathbf{A}(t')} |{\mathbf{r}}| 0 \rangle$ denotes the dipole matrix element for the bound-free transition.
Assuming that
\begin{eqnarray}
\langle \mathbf{r}| 0 \rangle=N_fe^{-\kappa r},
\end{eqnarray}
where $N_f$ is the normalization factor and $\kappa=\sqrt{2I_p}$.
The atom dipole matrix element can be written as
\begin{eqnarray}
\begin{aligned}
\textbf{d}_{\textrm{a}}(\mathbf{k})\sim\langle\mathbf{k}|\mathbf{r}|0\rangle\sim-i\frac{\mathbf{k}}{(\mathbf{k}^2+\kappa^2)^3},
\end{aligned}
\end{eqnarray}
where $\mathbf{k}=\mathbf{p}+\mathbf{A}(t')$ denotes the instantaneous momentum at the ionization time $t'$.
The final momentum distribution of the photoelectron is given by $|M_\mathbf{p}|^2$.

For a sufficiently high intensity and low frequency of the laser field, the temporal integration in Eq. (3)
can be evaluated by the saddle-point method \cite{Lewenstein1995,Figueira2002},
in which the solutions satisfy the stationary action equation
\begin{eqnarray}
\begin{aligned}
\frac{[\mathbf{p}+\mathbf{A}(t'_s)]^2}{2}+I_p=0.
\end{aligned}
\end{eqnarray}
Physically, Eq. (7) ensures the conservation of energy at the saddle-point time $t'_s$.
For $I_p>0$, Eq. (7) has only complex solutions $t'_s=t'_{sr}+it'_{si}$.
The real part $t'_{sr}$ has been interpreted as the ionization time
and the imaginary $t'_{si}$ can be considered as the tunneling time \cite{Lewenstein1994, Salieres2001}.
Using the solutions of Eq. (7), the transition amplitude of Eq. (3) can be rewritten as
\begin{eqnarray}
\begin{aligned}
M_\mathbf{p}=-i\sum_s G(\mathbf{p},t'_s)=-i\sum_s F(\mathbf{p},t'_s) e^{iS(\mathbf{p},t'_s)},
\end{aligned}
\end{eqnarray}
where $F(\mathbf{p},t'_s)=\sqrt{\frac{2\pi i}{\partial^2 S(\mathbf{p},t'_s)/ \partial {t'_s}^2}}{\mathbf{E}(t'_s)}\cdot\textbf{d}_{\textrm{a}}[\mathbf{p}+\mathbf{A}(t'_s)]$.
The index $s$ runs over the relevant saddle points, which are also termed as electron trajectories or quantum orbits.
In quantum mechanics, Eq. (8) corresponds to the coherent superposition of contributions from all relevant quantum orbits that lead to the same asymptotic momentum $\mathbf{p}$.

In our simulations, these two components of the orthogonally polarized 2D laser field are polarized along the $x$ axis and the $z$ axis, respectively.
Therefore, Eq. (7) can be written as
\begin{eqnarray}
\begin{aligned}
\frac{[p_x+A_x(t'_s)]^2}{2}+\frac{[p_z+A_z(t'_s)]^2}{2}+\frac{p_y^2}{2}+I_p=0.
\end{aligned}
\end{eqnarray}
Here, the terms $A_x(t)$ and $A_z(t)$ are the $x$ component and the $z$ component of the vector potential $\textbf{A}(t)$.
It should be mentioned that, the value of $p_y$ is equal to zero in classical mechanics.
However, in quantum mechanics, the value of $p_y$ can be nonzero with a Gauss-like distribution centered at $p_y=0$.
One can see that from Eq. (9), the term of ${p_y^2}/{2}$ is equivalent to increasing the ionization energy and then decreasing the ionization probability.
For this reason, the value of $p_y$ is set as zero in the following discussions.
For a complex time $t'_s$, the values of $A_x(t'_s)$ and $A_z(t'_s)$ also have complex forms.
We can divide $A_x(t'_s)$ and $A_z(t'_s)$ into the forms of
$A_x(t'_s)=A_x^\text{Re}(t'_s)+iA_x^\text{Im}(t'_s)$ and $A_z(t'_s)=A_z^\text{Re}(t'_s)+iA_z^\text{Im}(t'_s)$.
Then Eq. (9) can be decomposed into real and imaginary parts. The real part can be written as
\begin{eqnarray}
\begin{aligned}
&\frac{[p_x+A_x^\text{Re}(t'_s)]^2+[p_z+A_z^\text{Re}(t'_s)]^2}{2}\\
&-\frac{A_x^\text{Im}(t'_s)^2+A_z^\text{Im}(t'_s)^2}{2}+I_p=0,
\end{aligned}
\end{eqnarray}
and  the imaginary part  is
\begin{eqnarray}
\begin{aligned}
&[p_x+A_x^\text{Re}(t'_s)]A_x^\text{Im}(t'_s)+[p_z+A_z^\text{Re}(t'_s)]A_z^\text{Im}(t'_s)=0.
\end{aligned}
\end{eqnarray}
The above equation means that, because of the exist of the second field perpendicular to the fundamental field,
the real part of the instantaneous momentum of the electron in the tunneling process orthogonal to its imaginary part.
That is
\begin{eqnarray}
\begin{aligned}
\text{Re}[\mathbf{p}+\mathbf{A}(t'_s)]\cdot\text{Im}[\mathbf{p}+\mathbf{A}(t'_s)]=0.
\end{aligned}
\end{eqnarray}
Equation (12) shows that in 2D cases, the instantaneous momentum $\text{Re}[\mathbf{p}+\mathbf{A}(t)]$ at $t=t'_s$ can be nonzero.
This property is different from one-dimensional cases of linearly-polarized single-color laser fields where $\text{Re}[\mathbf{p}+\mathbf{A}(t'_s)]\equiv0$.
It arises from the fact that tunneling along one polarization direction of the 2D laser field indeed is influenced by the other polarization component
since contributions of these two components are strongly coupled together in tunneling, as Eqs. (10) and (11) show.
One can anticipate that the minor component of the 2D laser field has an important effect on tunneling even its laser amplitude is small in comparison with the main one.

On the other hand, Eq. (12) allows the form of solution of $\text{Re}[\mathbf{p}+\mathbf{A}(t'_s)]=0$,
which implies that the strong coupling between contributions of these two polarization components of the 2D laser field to tunneling partly decouples.
In the following, we call this solution of Eq. (12) with $\text{Re}[\mathbf{p}+\mathbf{A}(t'_s)]=0$ the partial-decoupling approximation
since it's form is somewhat similar to cases of two independent single-color laser fields, as to be discussed in Eq. (14).

For the case of OTC of Eq. (2), we have
\begin{equation}
\left\{
\begin{aligned}
A_x^\text{Re}(t'_s)=&A_1\text{cos}(\omega_1 t'_{sr})\text{cosh}(\omega_1 t'_{si}),\\
A_z^\text{Re}(t'_s)=&A_2\text{cos}(\omega_2 t'_{sr}+\phi)\text{cosh}(\omega_2 t'_{si}),\\
A_x^\text{Im}(t'_s)=&-A_1\text{sin}(\omega_1 t'_{sr})\text{sinh}(\omega_1 t'_{si}),\\
A_z^\text{Im}(t'_s)=&-A_2\text{sin}(\omega_2 t'_{sr}+\phi)\text{sinh}(\omega_2 t'_{si}).
\end{aligned}
\right.
\end{equation}
Here, $A_1={E_1}/{\omega_1}$ and $A_2={E_2}/{\omega_2}$ are the amplitudes of the vector potential of the fundamental and the second-harmonic fields, respectively.
The expressions will be used in our simulations.

\subsection{Partial-decoupling approximation}
In the partial-decoupling approximation, with assuming $\text{Re}[\mathbf{p}+\mathbf{A}(t'_s)]=0$,   Eq. (10) can be rewritten as
\begin{equation}
\left\{
\begin{aligned}
&p_x+A_x^\text{Re}(t'_s)=0,\\
&\frac{[A_x^\text{Im}(t'_s)]^2}{2}+\frac{[A_z^\text{Im}(t'_s)]^2}{2}=I_p,
\end{aligned}
\right.
\end{equation}
with the value of $p_z$  determined by the expression $p_z+A_z^\text{Re}(t'_s)=0$.
Equation (14) can be further understood as follows. In the partial-decoupling approximation, 2D tunneling is dominated by the main component of
the 2D laser field with large laser amplitude, while the minor component mainly influences the ionization probability.
With Eq. (14),  a set of saddle-point solutions $(\mathbf{p},t'_{s})$ can also be obtained.

For the case of OTC of Eq. (2) at $\phi=0$, we can get the relation between $p_x$ and $p_z$ in the partial-decoupling approximation. That is (see Appendix for details)
\begin{eqnarray}
\begin{aligned}
p_z=A_2(\frac{I_p}{U_{p1}+4\frac{U_{p2}}{U_{p1}}p_x^2}-\frac{p_x^2}{2U_{p1}}+1).
\end{aligned}
\end{eqnarray}
Here, $U_{p_1}={E_1^2}/{(4\omega_1^2)}$ and $U_{p_2}={E_2^2}/{(4\omega_2^2)}$ are the electron ponderomotive energies for the fundamental field and the second harmonic field, respectively.
Eq. (15) corresponds to a certain curve in a $p_x$ versus $p_z$ diagram, which generally gives the outside boundary of the bright (large amplitudes) parts of PMDs.
 We will discuss the predictions of Eq. (15) in detail later.

\subsection{Classical limit}
With assuming $I_p=0$ in Eq. (9), we arrive at the classical limit of 2D tunneling.
In this case, Eq. (9) at $p_y=0$ can be written as
\begin{equation}
\left\{
\begin{aligned}
&p_x+A_x(t')=0\\
&p_z+A_z(t')=0.
\end{aligned}
\right.
\end{equation}
Equation (16) has real solutions $t'$ only for some specific drift momenta $(p_x, p_z)$, which give a certain curve in momentum space. 
For these specific momenta, the electron leaving the nucleus at $t'$ has a zero instantaneous momentum. 
For the case of OTC of Eq. (2) at $\phi=0$, with Eq. (16), the relation between the drift momenta $p_x$ and $p_z$ in the classical limit can also be obtained. That is
\begin{eqnarray}
\begin{aligned}
p_z=A_2(-\frac{p_x^2}{2U_{p_1}}+1).
\end{aligned}
\end{eqnarray}
One can see that, the above expression follows from Eq. (15)  for $I_p=0$.

We mention that both of the partial-decoupling approximation and the classical limit imply $\text{Re}[\mathbf{p}+\mathbf{A}(t'_s)]=0$.
The second line of Eq. (14) imposes a third condition for the imaginary parts, while in the classical limit,
the reality of the saddle-point times $t'_s$ is assumed, which leaves two equations for one real unknown, viz. the time $t'$.
In Sec. III, we will compare the predictions of Eq. (15) and Eq. (17) with the local most-probable tunneling route in OTC, which we will discuss below.

\subsection{Most-probable tunneling routes}
To understand 2D tunneling, it is meaningful for studying one of the remarkable characteristics of PMD in 2D laser fields, i.e., the  local maximum of PMD. 
Such a local maximum corresponds to a certain saddle point $t'_s$ of Eq. (7) and this defines an under-the-barrier tunneling route.
In the following, we will call this route the local most-probable tunneling route or the simple one of the local most-probable route.

Before discussing the general form of the local most-probable tunneling routes, we first discuss a typical case of these routes, i.e.,
 the  most-probable tunneling route,
which is related to the maximum of PMD. Next, we discuss the condition for the most-probable route.
As the amplitudes of PMD are proportional to $e^{-b}$ \cite{Lewenstein1995}, where $b$ is the imaginary part $S^{\text{Im}}(\mathbf{p},t'_s)$ of the semiclassical action $S(\mathbf{p},t'_s)$ of Eq. (4),
the  most-probable tunneling route can be obtained with finding the minimum of $b$, which corresponds to the stationary points of 
$S^{\text{Im}}(\mathbf{p},t'_s)$ with respect to  $p_x$ and $p_z$. That is
\begin{eqnarray}
\begin{aligned}
\frac{\partial S^{\text{Im}}(\mathbf{p},t'_s)}{\partial p_x}=0, \frac{\partial S^{\text{Im}}(\mathbf{p},t'_s)}{\partial p_z}=0.
\end{aligned}
\end{eqnarray}

For the case of OTC of Eq. (2), we have (see Appendix for details)
\begin{eqnarray}
\begin{aligned}
\frac{\partial S^{\text{Im}}(\mathbf{p},t'_s)}{\partial p_x}&=p_x t'_{si}+\frac{A_1}{\omega_1}\text{cos}(\omega_1 t'_{sr})\text{sinh}(\omega_1 t'_{si})=0\\
\frac{\partial S^{\text{Im}}(\mathbf{p},t'_s)}{\partial p_z}&=p_z t'_{si}+\frac{A_2}{\omega_2}\text{cos}(\omega_2 t'_{sr}+\phi)\text{sinh}(\omega_2 t'_{si})=0.
\end{aligned}
\end{eqnarray}
According to Eq. (19), when the amplitude of the photoelectron arrives at the maximum at the saddle point $t'_s$, the values of $p_x$ and $p_z$ agree with
$p_x=-\frac{1}{t'_{si}}\frac{A_1}{\omega_1}\text{cos}(\omega_1 t'_{sr})\text{sinh}(\omega_1 t'_{si})$, and
$p_z=-\frac{1}{t'_{si}}\frac{A_2}{\omega_2}\text{cos}(\omega_2 t'_{sr}+\phi)\text{sinh}(\omega_2 t'_{si})$.

\subsection{Local most-probable tunneling routes}
Next, we discuss the local most-probable tunneling route associated with the regions with the local maximal amplitudes
in PMD.

With rotating the coordinate axis from the experimental frame to that of the instantaneous combined field
and differentiating  $S^{\text{Im}}(\mathbf{p},t'_s)$ in the new coordinate system,
analytical expressions associated with  transcendental equations
for conditions of the local most-probable tunneling routes have  been obtained
for elliptical \cite{Luo2019} and OTC \cite{XieW2019} laser fields.
Here, we address this issue from the perspective of classic-quantum correspondence and find a general condition for these local routes.
We assume that saddle points of Eq. (7) have been obtained.

To find the condition,
we first analyze the  position of the tunnel exit at the time $t'_{sr}$ (the real part of the saddle-point time $t'_s$),
which is considered as the time at which the electron tunnels out of the barrier. That is \cite{Beckeroribit,TMYan2012}
\begin{eqnarray}
\begin{aligned}
\mathbf{r}_0(\mathbf{p},t'_{sr})=\int_{t'_s}^{t'_{sr}} [\mathbf{p}+\mathbf{A}(t'')] dt''.
\end{aligned}
\end{eqnarray}
The position vector $\mathbf{r}_0(\mathbf{p},t'_{sr})$ is also complex with the form of
($x^{\text{Re}}_0(\mathbf{p},t'_{sr}),z^{\text{Re}}_0(\mathbf{p},t'_{sr})$)+i($x^{\text{Im}}_0(\mathbf{p},t'_{sr}),z^{\text{Im}}_0(\mathbf{p},t'_{sr})$).
For most of saddle points, the imaginary part ($x^{\text{Im}}_0(\mathbf{p},t'_{sr}),z^{\text{Im}}_0(\mathbf{p},t'_{sr})$) of the exit-position vector is  nonzero.

From a semiclassical view of point, as the electron tunnels out of the barrier,
it begins to behave classically.
The classical behavior and the quantum behavior of the electron connect at the tunnel exit.
This implies that the smaller the imaginary part  of the exit position is, the nearer to the classical one
the electron is at the tunnel exit. We therefore conclude that electron trajectories associated with smaller imaginary parts of the exit position have
larger amplitudes \cite{chenchao2019}. This is the semiclassical condition which we are finding for this local most-probable route. With defining
\begin{eqnarray}
\begin{aligned}
\text{r}^\text{Im}_0(\mathbf{p},t'_{sr})=&\sqrt{[x^\text{Im}_{0}(\mathbf{p},t'_{sr})]^2+[z^\text{Im}_{0}(\mathbf{p},t'_{sr})]^2},
\end{aligned}
\end{eqnarray}
the condition for the local most-probable route of $(\textbf{p}_x,\textbf{p}_z)$ can be written as
\begin{eqnarray}
\begin{aligned}
\text{Min}[\text{r}^\text{Im}_0(\mathbf{p},t'_{sr})]\equiv\text{Min}[\text{r}^\text{Im}_0(\mathbf{p}_x,\mathbf{p}_z,t'_{sr})].
\end{aligned}
\end{eqnarray}
Here, the sign ``Min'' denotes the minimal value of the relevant function. We will also use the sign in Fig. 3.
 Numerically, with the above expressions, evaluations on this local route, (e.g., for a certain value of ${p}_x$, to find the value of ${p}_z$
such that the PMDs at $({p}_x,{p}_z)$ have the local maximal amplitudes),
convert into calculating the exit position $\mathbf{r}_0(\mathbf{p},t'_{sr})$ of Eq. (20), and the condition of Eq. (22)
can be conveniently treated with numerical analysis on minimal values of Eq. (21).
Analytically, the extreme values of $\text{r}^\text{Im}_0(\mathbf{p},t'_{sr})$ are usually the arrest points or non-differentiable points of the function $\partial\text{r}^\text{Im}_0(\mathbf{p},t'_{sr})/\partial{p_z}$.
As we will show in the paper,  the condition of Eq. (22) is applicable for 2D laser fields with different forms.
It gives a good description for the local maximal amplitudes in PMDs.
Specific momentum pairs ($p_x,p_z$) agreeing with Eq. (22)  give a curve in a $p_x$ versus $p_z$ diagram,
which characterizes the PMD of orthogonally polarized 2D laser fields and can be directly compared with the corresponding curves of the partial-decoupling approximation
and the classical limit.

One can expect that  with the condition of
\begin{eqnarray}
\begin{aligned}
\text{r}^\text{Im}_0(\mathbf{p},t'_{sr})\rightarrow0,
\end{aligned}
\end{eqnarray}
relevant routes predicted by Eq. (22)   will show larger amplitudes.
In addition, for smaller values of imaginary time $t'_{si}$, Eq. (23) will be easier to satisfy.

For the case of OTC of Eq. (2), the real and imaginary parts of the exit position can be expressed as
\begin{eqnarray}
\left\{
\begin{aligned}
x^{\text{Re}}_0(\mathbf{p},t'_{sr})=&\frac{A_1}{{\omega_1}}\text{sin}(\omega_1 t'_{sr})[1-\text{cosh}(\omega_1 t'_{si})]\\
z^{\text{Re}}_0(\mathbf{p},t'_{sr})=&\frac{A_2}{{\omega_2}}\text{sin}(\omega_2 t'_{sr}+\phi)[1-\text{cosh}(\omega_2 t'_{si})]
\end{aligned}
\right.
\end{eqnarray}
and
\begin{eqnarray}
\left\{
\begin{aligned}
x^{\text{Im}}_0(\mathbf{p},t'_{sr})=&-p_x t'_{si}-\frac{A_1}{{\omega_1}}\text{cos}(\omega_1 t'_{sr})\text{sinh}(\omega_1 t'_{si})\\
z^{\text{Im}}_0(\mathbf{p},t'_{sr})=&-p_z t'_{si}-\frac{A_2}{{\omega_2}}\text{cos}(\omega_2 t'_{sr}+\phi)\text{sinh}(\omega_2 t'_{si}).
\end{aligned}
\right.
\end{eqnarray}
One can observe that there is a simple relation between Eqs. (25) and (19).

\subsection{Comparisons with one-dimensional tunneling}
For comparison, here, we simply discuss cases of one-dimensional (1D) tunneling such as atoms exposed to strong linearly-polarized single-color laser fields.
With assuming $p_y=p_z=0$ for the reason as discussed below Eq. (9), the saddle-point equation for 1D cases is similar to Eq. (14) and can be written as
\begin{equation}
\left\{
\begin{aligned}
&p_x+A_x^\text{Re}(t'_s)=0,\\
&\frac{[A_x^\text{Im}(t'_s)]^2}{2}=I_p.
\end{aligned}
\right.
\end{equation}
The classical limit corresponding to Eq. (16) becomes
\begin{equation}
\begin{aligned}
p_x+A_x(t')=0,
\end{aligned}
\end{equation}
and the condition for the most-probable tunneling route corresponding to Eq. (18) becomes
\begin{eqnarray}
\begin{aligned}
\frac{\partial S^{\text{Im}}(\mathbf{p},t'_s)}{\partial p_x}&=p_x t'_{si}+\frac{A_1}{\omega_1}\text{cos}(\omega_1 t'_{sr})\text{sinh}(\omega_1 t'_{si})=0.
\end{aligned}
\end{eqnarray}
Due to the disappearance of the second field perpendicular to the fundamental field,
in 1D cases, the local maximal amplitudes corresponding to different values of $p_x$ in PMDs always appear along the axis of $p_z=0$,
in agreement with the classical predictions. So Eq. (22) is no longer needed here. However, Eq. (23) still works.
As we will discuss in Fig. 3, the predictions of Eq. (23) in 1D cases agree with the intuitive presumption that tunneling prefers a small imaginary time.
Nevertheless, this presumption is not applicable for 2D tunneling when Eq. (23) still gives a good description for the preference of tunneling.

It should also be stressed that in 1D the solutions of $p_x+A_x(t)=0$ are real for any $|p_x|<|\max(A_x(t))|$.
In contrast, in 2D, even in the classical limit there are no real solutions for general $p_x$ and $p_z$, because the two equations $p_x+A_x(t)=0$ and $p_z+A_z(t)=0$
will not both be solved simultaneously by the same solution $t$. Hence, in general, the solutions will be complex, except for certain special values of $p_x$ and $p_z$.
However, nonzero imaginary parts lead to decreased ionization rates.
Hence, for the maxima of the momentum-dependent rate, we attempt to find curves in momentum space along which the saddle-points are real.
A similar reasoning is invoked to find the optimal phase for HHG by an OTC field in  \cite{Milosevi2019}.
These special values of $p_x$ and $p_z$ can be obtained with finding the time $t$ which agrees with the equation $p_x+A_x(t)=0$ for a specific value of $p_x$ with $|p_x|<|\max(A_x(t))|$.
Then the corresponding value of $p_z$ can be obtained with the equation $p_z+A_z(t)=0$ at this time $t$. In addition, for some special forms of 2D laser fields,
such as OTC laser fields and circular laser fields explored in the paper,
a simple relation between $p_x$ and $p_z$ for these special values can also be obtained, as indicated by Eq. (17) and Eq. (33) in the paper.

\subsection{Transition towards classical limit}
As discussed above, with Eq. (22), one can obtain the local most-probable routes in 2D tunneling, the amplitudes of which are scaled with Eq. (23).
Another question one can concern is when the local most-probable routes will approach the classical limit.
To answer this question, we consider a pair of drift momenta $(p^q_x,p^q_z)$ associated with the local most-probable route and corresponding to the saddle-point time $t'_s=t'_{sr}+it'_{si}$.
For the case of OTC of Eq. (2), according to Eq. (25), they can be written as
\begin{eqnarray}
\left\{
\begin{aligned}
&p^q_x=-A_1\text{cos}(\omega_1 t'_{sr})\frac{\text{sinh}(\omega_1 t'_{si})}{\omega_1 t'_{si}}-\frac{x^{\text{Im}}_0(\mathbf{p},t'_{sr})}{t'_{si}},\\
&p^q_z=-A_2\text{cos}(\omega_2 t'_{sr}+\phi)\frac{\text{sinh}(\omega_2 t'_{si})}{\omega_2 t'_{si}}-\frac{z^{\text{Im}}_0(\mathbf{p},t'_{sr})}{t'_{si}}.
\end{aligned}
\right.
\end{eqnarray}
Classically, at the corresponding ionization time $t'_{sr}$, the electron has the drift momenta
\begin{eqnarray}
\left\{
\begin{aligned}
&p^c_x=-A_1\text{cos}(\omega_1 t'_{sr}),\\
&p^c_z=-A_2\text{cos}(\omega_2 t'_{sr}+\phi).
\end{aligned}
\right.
\end{eqnarray}
The approach of quantum predictions towards the classical limit implies
\begin{equation}
\left\{
\begin{aligned}
&p^q_x-p^c_x \rightarrow0,\\
&p^q_z-p^c_z \rightarrow0.
\end{aligned}
\right.
\end{equation}
One can see that, with the conditions of Eq. (23) and $\lim\limits_{t'_{si}\to0}\frac{\text{sinh}(\omega_{1,2} t'_{si})}{\omega_{1,2} t'_{si}}=1$,
the smaller the values of $t'_{si}$ and $\text{r}^{\text{Im}}_0(\mathbf{p},t'_{sr})$ are, the closer the local most-probable route will be to the classical limit.
Predictably, with the decrease of the Keldysh parameter for both these two components of 2D laser fields, Eq. (31) is possible to fulfill as to be discussed in Sec. III. C.

\section{Cases of OTC with zero time delay}

\begin{figure}[t]
\begin{center}
\rotatebox{0}{\resizebox *{7.0cm}{8cm} {\includegraphics {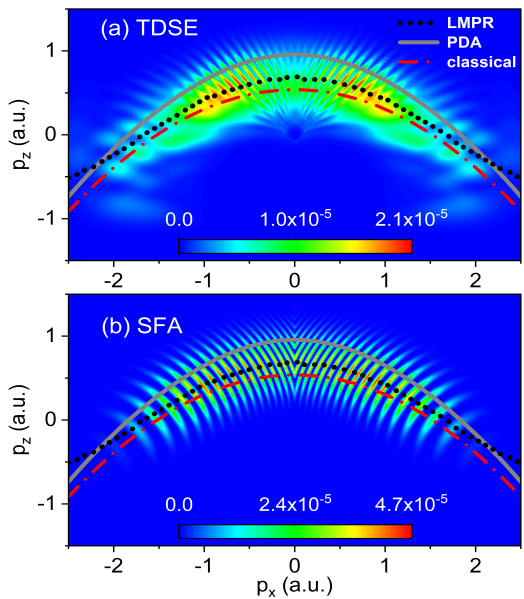}}}
\end{center}
\caption
{(Color online) PMDs of model He atom in the OTC laser field of $\phi=0$, obtained with TDSE (a) and SFA (b).
The predictions of Eq. (22) for the local most-probable route (LMPR) (black dotted), Eq. (15) for the partial-decoupling approximation (PDA) (gray solid),
and Eq. (17) for the classical limit (red dashed-dotted) are plotted here with different curves.
}  \label{Fig. 1}
\end{figure}
\begin{figure}[t]
\begin{center}
\rotatebox{0}{\resizebox *{7cm}{7cm} {\includegraphics {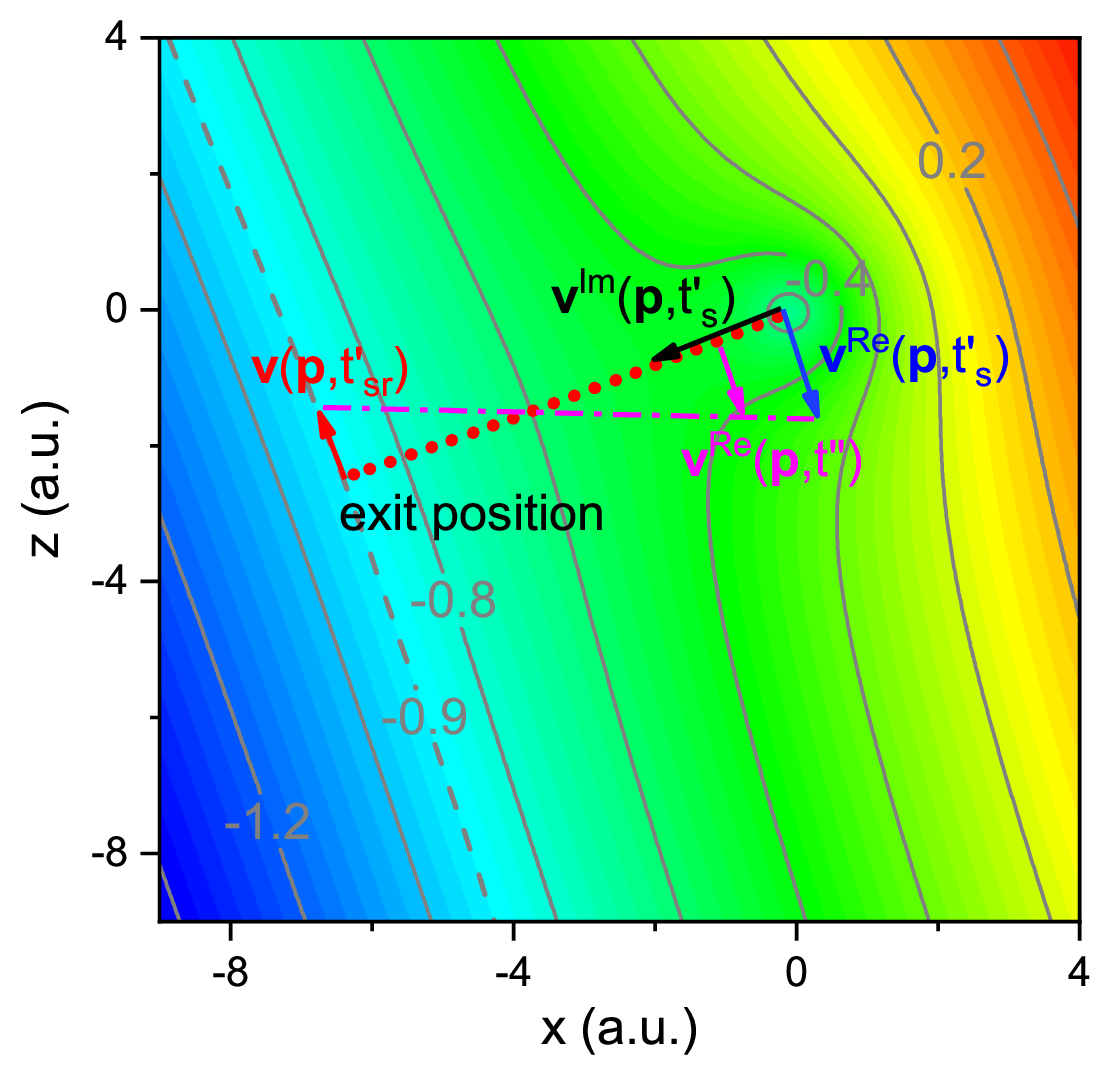}}}
\end{center}
\caption
{(Color online) The schematic diagram of the 2D tunneling process under the potential barrier of model He atom in the OTC field of $\phi=0$. The color code of the map denotes the potential created by the Coulomb field and the electric field at a certain time $t$.
The contour lines with the numbers (in units of a.u.) show the corresponding values of the potential. The dashed contour line shows the potential at the value of -0.9 a.u..
The blue-solid and black-solid arrows show the real and the imaginary parts of the instantaneous velocity at the time $t'_s$, respectively.
The magenta-solid arrow shows the real part of the instantaneous velocity in the tunneling process,
and its magnitudes at different times are shown with the magenta dashed-dotted line.
The red-dotted line represents the tunneling direction.
The red-solid arrow shows the final velocity at the time $t'_{sr}.$
}  \label{Fig. 2}
\end{figure}
\begin{figure*}[t]
\begin{center}
\rotatebox{0}{\resizebox *{16.8cm}{7.5cm} {\includegraphics {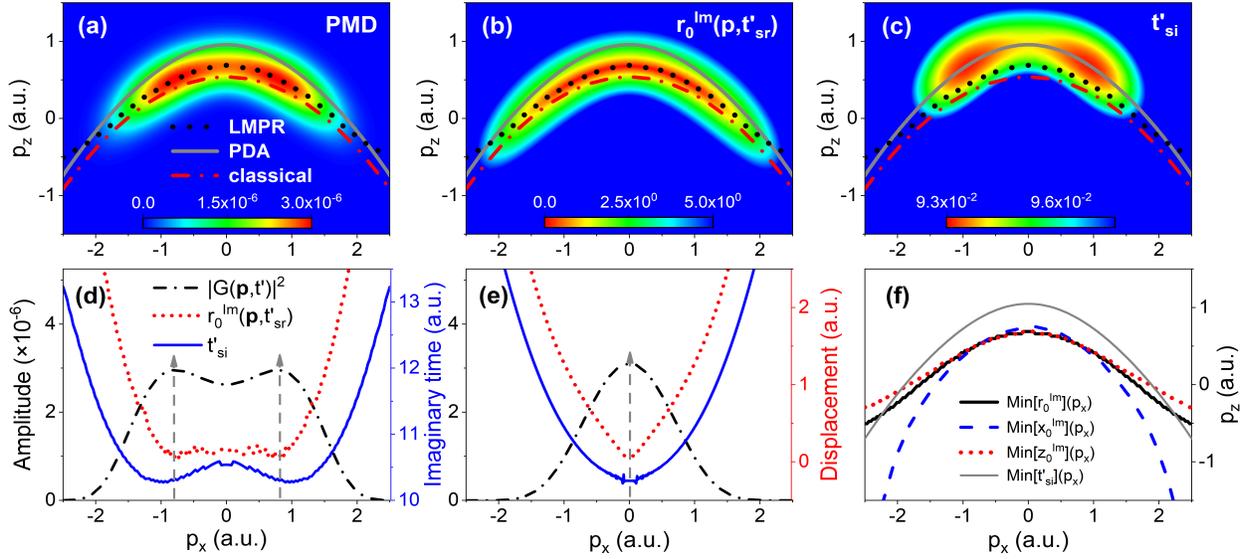}}}
\end{center}
\caption
{(Color online) PMD (a), distributions of imaginary position $\text{r}_0^{\text{Im}}(\mathbf{p}, t'_{sr})$ (b)
and imaginary time $t'_{si}$ (c) for model He atom in the OTC laser field of $\phi=0$,
obtained with SFA and calculated for a half laser cycle of the fundamental field.
The predictions of Eq. (22) for the local most-probable route (LMPR) (black dotted), Eq. (15) for the partial-decoupling approximation (PDA) (gray solid),
and Eq. (17) for the classical limit (red dashed-dotted) are plotted in (a)-(c) with different curves.
In (d), amplitudes $\mid G(\mathbf{p},t'_s)\mid^2$ (black dashed-dotted), imaginary position $\text{r}_0^{\text{Im}}(\mathbf{p},t'_{sr})$ (red dotted)
and imaginary time $t'_{si}$ (blue solid), associated with LMPR, are plotted as functions of $p_x$.
Results in (e) are similar to (d), but assuming that the second harmonic field disappears.
In (d) and (e), each color curve uses the vertical coordinate axis of the same color.
In (f), we show specific momenta $p_z$ as functions of $p_x$, agreeing with the conditions of Eq. (22) (black solid),
$\text{Min}[\text{x}^\text{Im}_0]$ (blue dashed), $\text{Min}[\text{z}^\text{Im}_0]$ (red dotted), and $\text{Min}[{t'_{si}}]$ (gray thin-solid).
}  \label{Fig. 3}
\end{figure*}

\subsection{Local maximal amplitudes in PMDs}
In Fig. 1, we show the PMDs of TDSE and SFA with saddle-point method for the model He atom in the OTC laser field with zero time delay.
The TDSE results in Fig. 1(a) show a fan-like structure with bright parts having large amplitudes located in the middle of the distribution.
These characteristics are basically reproduced by the SFA, as seen in Fig. 1(b).
Both of TDSE and SFA results show clear radial interference fringes, which can be attributed to the intra-cycle interference of electron trajectories (see Appendix for details).

The predictions of Eqs. (15), (17) and (22) for specific momentum pairs ($p_x,p_z$) are plotted in Fig. 1 with different curves.
Indeed, one can observe that the curve of Eq. (22), associated with the local most-probable tunneling routes,
goes through the bright parts (with large amplitudes) of the distributions in Figs. 1(a) and 1(b), suggesting the applicability of Eq. (22).
On the other hand, as the predictions of Eq. (15) for the partial-decoupling approximation give the outside boundaries of the bright parts of the distributions,
the predictions of Eq. (17) for the classical limit give the inside ones. The theoretical curves characterize these PMDs related to 2D tunneling.

Since the predictions of Eq. (22), arising from SFA with the saddle-point theory, for the local maximal amplitudes in PMDs agree with the TDSE ones,
next, we make a detailed analysis on the implications of Eq. (22).

\subsection{Analyses on local most-probable routes}

To understand the 2D tunneling process, in Fig. 2, we present a sketch of this process  in OTC of $\phi=0$, which, according to the saddle-point theory,
is described as the complex-time evolution of the electronic wave packet under the barrier.
We assume that the electronic wave packet was at the origin of the nucleus at the time $t'_s$.
Just below Eq. (12), it has been mentioned that at the time $t'_s$, the real (the blue solid arrow)
and imaginary (the black solid arrow) parts of the instantaneous velocity $\mathbf{v}(\mathbf{p},t'_s)$ are perpendicular to each other.
The imaginary part of the instantaneous velocity is along the tunneling direction where the barrier changes the fastest (the red dotted curve).
Then, with the evolution of the time from $t'_s$ to $t'_{sr}$,
the imaginary part of the instantaneous velocity $\mathbf{v}^\text{Im}(\mathbf{p},t'')$ changes from $\mathbf{A}(t'_s)$ to zero,
and the integral over the imaginary velocity along the imaginary time axis generates a real displacement $\mathbf{r}_0^\text{Re}(\mathbf{p},t'_{sr})$ along the tunneling direction.
At the time $t'_{sr}$, one can see that the electronic wave packet tunnels out of the barrier and has only a real velocity $\mathbf{v}(\mathbf{p},t'_{sr})$.
On the other hand, in the middle time $t''$, the real part of the velocity $\mathbf{v}^\text{Re}(\mathbf{p},t'')$ (the magenta solid arrow) is always perpendicular to the tunneling direction,
and its magnitudes at different times are shown with the magenta dashed-dotted line in Fig. 2.
For the imaginary-time evolution, the velocity $\mathbf{v}^\text{Re}(\mathbf{p},t'')$ does not contribute to the displacement in the real space,
however it can generate an imaginary displacement $\mathbf{r}_0^\text{Im}(\mathbf{p},t'_{sr})$.
This is the origin of the term defined with Eq. (21).

In Sec. II. F, we have conjectured that the imaginary displacement $\text{r}_0^\text{Im}(\mathbf{p},t'_{sr})$ dominates the tunneling probability.
We validate this conjecture in Fig. 3, where we plot SFA results with the ionization events occurring only in a half optical cycle
of the fundamental field of OTC for clarity.

Results of PMD are presented in Fig. 3(a),  where the intra-cycle interference is absent due to simulations with a half optical cycle.
The predictions of Eq. (22) for the local most-probable route (the black dotted line), Eq. (15) for the partial-decoupling approximation (the gray solid line)
and Eq. (17) for the classical limit (the red dashed-dotted line) are plotted here with diverse curves.
One can see that, the black dotted line coincides with the brightest parts (corresponding to large amplitudes here) of the distribution,
while the gray and the red curves give the outside and the inside boundaries of the bright parts of the distributions, as in Fig. 1.
In Fig. 3(b), we show the distribution of $\text{r}_0^{\text{Im}}(\mathbf{p},t'_{sr})$. The black dotted lines of  Eq. (22) also just
goes through the bright parts (corresponding to small values here) of the distribution, while the behaviors of the gray and red curves here are also similar to those in Fig. 3(a).

In Fig. 3(c), we plot the distribution of the imaginary time $t'_{si}$.
One can expect that electron trajectories ($\mathbf{p},t'_s$) with smaller imaginary times $t'_{si}$ will have larger amplitudes,
since at the limit of $t'_{si}=0$, relevant quantum trajectories will return to the classical case.
However, it is not the case seen in Fig. 3(c), where the predictions of Eq. (22) for the local most-probable route
do not agree with the brightest parts (corresponding to small values here) of the distribution.
Instead, the predictions of Eq. (15) for the partial-decoupling approximation go through the brightest parts,
implying that electron trajectories of the partial-decoupling approximation have the smallest imaginary part of the saddle-point time $t'_s=t'_{sr}+it'_{si}$,
in comparison with those of the local most-probable and the classical ones.

To obtain more insights into the properties of the local most-probable routes defined with Eq. (22), in Fig. 3(d),
we plot the curves of amplitude $\mid G(\mathbf{p},t'_s)\mid^2$ (the black dashed-dotted curve),
imaginary position $\text{r}_0^{\text{Im}}(\mathbf{p},t'_{sr})$ (the red dotted curve)
and imaginary time $t'_{si}$ (the blue solid curve) of the local most-probable routes of Eq. (22), as functions of
$p_x$. These two peaks of the amplitude curve  are marked by the gray dashed arrows. It can be seen that
the positions of these two peaks  agree with the minimum locations of the imaginary-position curve, but differ somewhat from those of the imaginary-time curve.

This situation is different from the tunneling process in linearly-polarized single-color laser fields, as shown in Fig. 3(e).
For the case of one-dimensional tunneling in Fig. 3(e), the amplitude curve of $\mid G(\mathbf{p},t'_s)\mid^2$ shows a single peak around $p_x=0$
at which the curves of the imaginary time and position show the minima simultaneously.
More specifically, results in Fig. 3(e) tell that for one-dimensional tunneling, as the imaginary time gets smaller,
the imaginary position also does so and the tunneling amplitude becomes larger.
However, in the OTC field, this conclusion is no longer applicable.
Because of the participation of tunneling along the $z$ axis relating to the second harmonic field,
the tunneling amplitude is not maximal when the imaginary time $t'_{si}$ is minimal.

We therefore judge that the imaginary  position $\text{r}_0^{\text{Im}}(\mathbf{p},t'_{sr})$
is the critical quantity for influencing the probabilities of the corresponding tunneling event.
Tunneling prefers electron trajectories with the condition of Eq. (23), i.e., $\text{r}_0^{\text{Im}}(\mathbf{p},t'_{sr})\rightarrow0$.

To further understand the condition of Eq. (22), in Fig. 3(f),
we plot curves of specific momentum pairs ($p_x,p_z$) which agree with the conditions of  Eq. (22) (the black solid curve),
$\text{Min}[\text{x}^\text{Im}_0]$  (the blue dashed curve),
$\text{Min}[\text{z}^\text{Im}_0]$  (the red dotted curve),
and $\text{Min}[{t'_{si}}]$ (the gray thin-solid curve).
Specifically, for a certain value of $p_x$, we find the corresponding values of $p_z$,
which minimize the values of imaginary-position functions $\text{r}^\text{Im}_0$, $\text{x}^\text{Im}_0$,
$\text{z}^\text{Im}_0$, and the imaginary time $t'_{si}$, respectively.
One can observe that the curve of $\text{r}_0^{\text{Im}}$, corresponding to the local most-probable tunneling route,
is nearer to that of $\text{z}_0^{\text{Im}}$, in comparison with the curve of $\text{x}_0^{\text{Im}}$.
The comparisons suggest that in OTC laser fields, it is the tunneling motion along the polarization direction of the second harmonic field
which plays a dominating role in the tunneling amplitudes.
The curve with the minimal imaginary time $t'_{si}$ deviates remarkably from all of these imaginary-position curves,
suggesting the inapplicability of simply relating the tunneling probabilities to $t'_{si}$.
This point has been discussed in Fig. 3(c).

\subsection{Roles of Keldysh parameter}
In the above discussions, we have explored some characteristics of 2D tunneling in the OTC laser field,
including the local most-probable route and its condition, with the laser parameters of $\phi=0$, $I_1=5\times 10^{14} \text{W/cm}^2$, $\lambda_1=800$ nm and $\varepsilon=0.5$.

When the value of the Keldysh parameter $\gamma=\sqrt{I_p/2U_p}$ decreases, the tunneling event is easier to occur.
In this case, one can also expect that the behavior of the tunneling electron born at the time $t'_{sr}$ will be closer to the classical one.
In the following, based on simulations of TDSE and SFA, we further study the local most-probable route, which can be considered as the quantum counterpart of the classical limit,
for OTC laser field with other laser parameters of $I_1=1\times 10^{14} \text{W/cm}^2$, $\lambda_1=1400$ nm, $\varepsilon=1$ or $\varepsilon=2$,
corresponding to relatively stronger second-harmonic fields and smaller values of $\gamma$.
Relevant results are presented in Fig. 4.

\begin{figure}[t]
\begin{center}
\rotatebox{0}{\resizebox *{8.7cm}{8.7cm} {\includegraphics {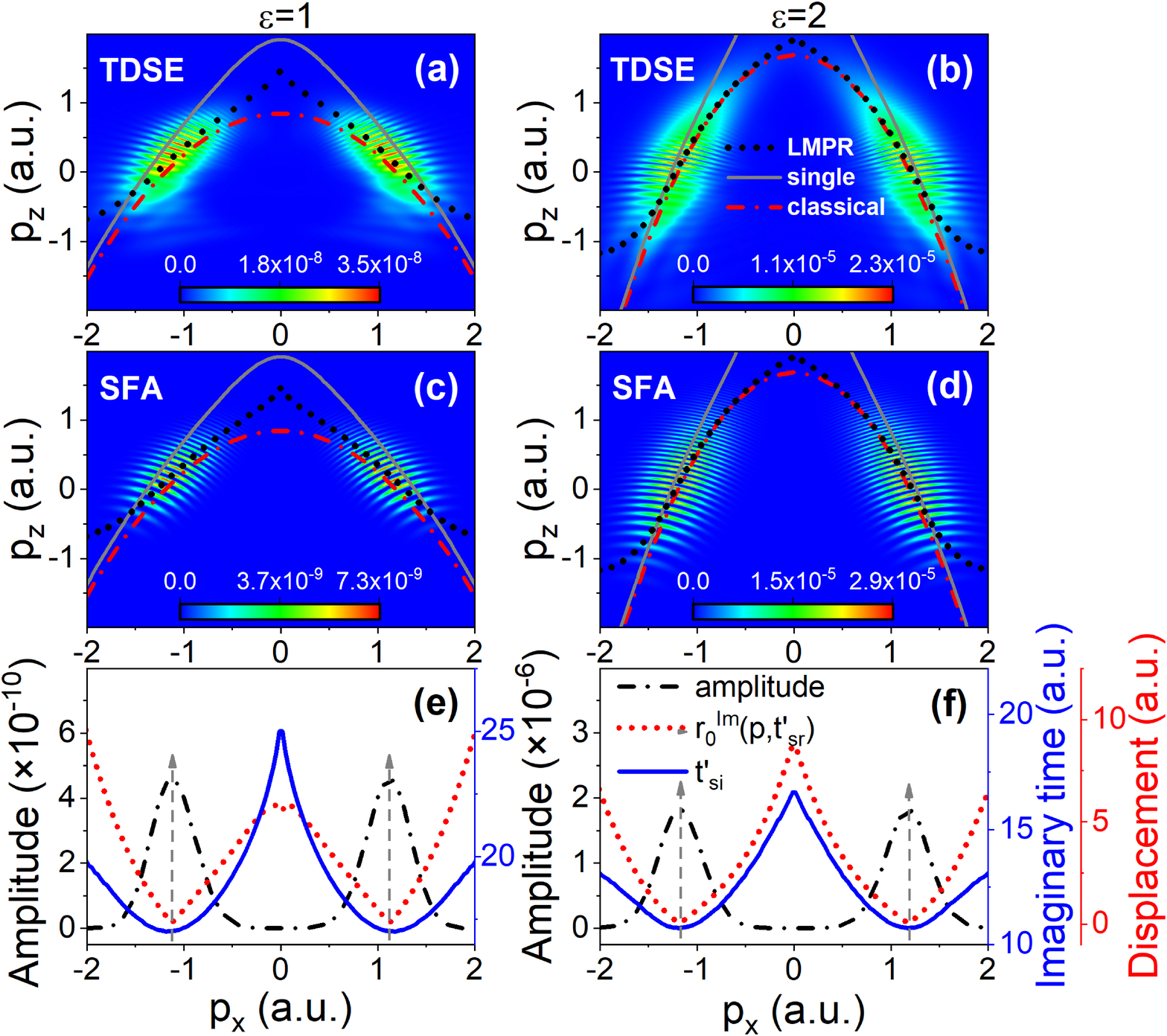}}}
\end{center}
\caption
{(Color online)
PMDs of model He atom  in OTC laser fields  of $\phi=0$, obtained with TDSE (a,b) and SFA (c,d).
The laser parameters used are $I_1=1\times10^{14}\text{W/cm}^2$ and $\lambda_1=1400$ nm for the fundamental field, with $\varepsilon=1$ in (a,c,e) and $\varepsilon=2$ in (b,d,f).
The predictions of Eq. (22) for the local most-probable route (LMPR) (black dotted), Eq. (15) for the partial-decoupling approximation (PDA) (gray solid), and Eq. (17)
for the classical limit (red dashed-dotted) are plotted  in (a)-(d) with different curves. Results in (e) and (f) are similar to those in Fig. 3(d), but for present OTC laser parameters.
} \label{Fig. 4}
\end{figure}

Firstly, it can be seen from Figs. 4(a) and 4(b) that, PMDs of TDSE are divided into two parts which are symmetric about $p_x=0$
and the bright parts of PMDs with large amplitudes shift towards larger values of $p_x$, in comparison with results in Fig. 1(a).
These characteristics are in good agreement with the SFA predictions in Figs. 4(c) and 4(d).
Secondly, for the cases in the left column of Fig. 4 with $\varepsilon=1$,
the predictions of Eq. (22) also go through the bright parts of these distributions, while those of Eqs. (15) and (17)
give the outside and the inside boundaries of these bright parts.
In particular, the curve of Eq. (17) for the classical limit becomes closer to that of Eq. (22) for the local most-probable route,
suggesting a closer correspondence between classic and quantum.
This tendency becomes more remarkable in the right column of Fig. 4 with $\varepsilon=2$ corresponding to a smaller value of $\gamma$ than  $\varepsilon=1$,
where these curves of Eqs. (17) and (22) are almost coincident with each other in the main parts of the distributions.
These phenomena suggest that for full small values of $\gamma$, Eq. (29) can be well satisfied and the electron born at the time $t'_{sr}$ becomes classical-like.

Comparisons between these curves in Figs. 4(e) and 4(f) also show that the peak locations of amplitude curves
agree with the minimum locations of imaginary-position $\text{r}_0^{\text{Im}}$ curves.
In particular, for the case of $\varepsilon=2$ in Fig. 4(f),
the minimum locations of imaginary-time curves also agree with those of $r_0^{\text{Im}}$ curves,
implying the recovery of the intuitive preassumption that tunneling is easier to occur for smaller imaginary times.
Combing the results in Fig. 3 and Fig. 4, we can also conclude that the disagreement of minimum locations between imaginary time and position revealed in Fig. 3(d) is
a signature of the deviation between quantum and classical predictions for the local most-probable route in strong-field 2D tunneling.

\section{Cases of circular laser fields}

To validate our conclusion, we also perform simulations for cases of the circularly-polarized laser field, which has the following form
\begin{eqnarray}
\mathbf{E}(t)=f(t)\frac{E}{\sqrt{1+\epsilon^2}}[{\sin(\omega t)\mathbf{e}_x}+{\epsilon\cos(\omega t)\mathbf{e}_z}].
\end{eqnarray}
Here, $E$ is the laser amplitude corresponding to the peak intensity $I$ and $\epsilon$ is the ellipticity with $\epsilon=1$ for circularity.
The classical limit (corresponding to Eq. (17) of OTC) in circular cases can be written as
\begin{eqnarray}
\begin{aligned}
p_x^2+p_z^2=4U_p,
\end{aligned}
\end{eqnarray}
and the partial-decoupling approximation (corresponding to Eq. (15) of OTC) is
\begin{eqnarray}
\begin{aligned}
p_x^2+p_z^2=4U_p+2I_p.
\end{aligned}
\end{eqnarray}
Here, $U_p=E^2/4\omega^2$. The imaginary parts of the exit position for electron trajectories can be expressed as
\begin{eqnarray}
\left\{
\begin{aligned}
x^{\text{Im}}_0(\mathbf{p},t'_{sr})=&-p_x t'_{si}-\frac{A}{{\omega}}\text{cos}(\omega t'_{sr})\text{sinh}(\omega t'_{si})\\
z^{\text{Im}}_0(\mathbf{p},t'_{sr})=&-p_z t'_{si}-\frac{A}{{\omega}}\text{sin}(\omega t'_{sr})\text{sinh}(\omega t'_{si}),
\end{aligned}
\right.
\end{eqnarray}
and the form of Eq. (22) for the local most-probable route is unchanged.

\begin{figure}[t]
\begin{center}
\rotatebox{0}{\resizebox *{8.5cm}{8.5cm} {\includegraphics {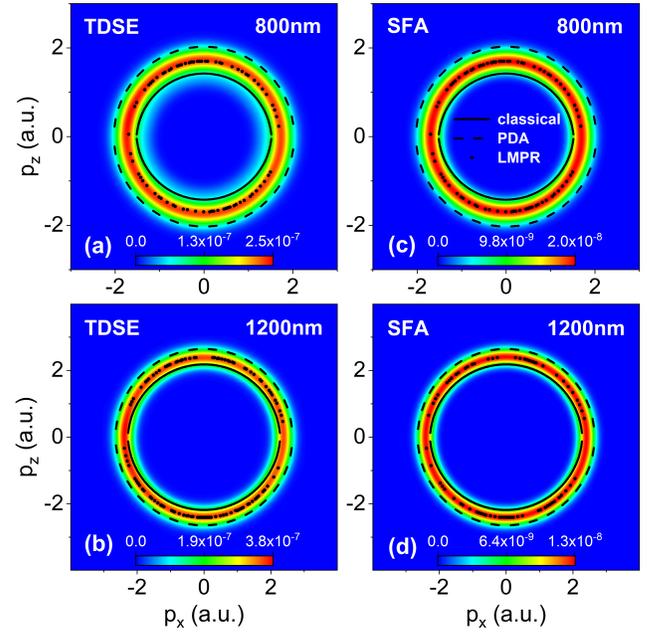}}}
\end{center}
\caption
{(Color online) PMDs of model He atom in  circular laser fields, obtained with TDSE (a,b) and SFA (c,d).
The peak intensity of the circular laser field is $I=5\times10^{14}\text{W/cm}^{2}$
with the wavelength $\lambda=800$ nm (the first row) and $\lambda=1200$ nm (second).
The predictions of Eq. (22) for the local most-probable route (LMPR) (dotted),  Eq. (33) for the classical limit (solid),
and Eq. (34) for the partial-decoupling approximation (PDA) (dashed) are plotted here with different curves.
} \label{Fig. 5}
\end{figure}

In Fig. 5, we plot  PMDs of TDSE  and SFA simulations for circular laser fields with the laser intensity of $I=5\times10^{14}\text{W/cm}^{2}$
and the wavelength of $\lambda=800$ nm (the first row) and $\lambda=1200$ nm (the second row), respectively.
Also shown are the predictions of the classical limit of Eq. (33) (the solid curve), the partial-decoupling approximation of Eq. (34) (the dashed curve)
and the local most-probable route defined with Eq. (22) (the dotted curve).

It can be seen from Fig. 5 that, the PMDs present  a ring structure and the bright parts of these distributions,
 corresponding to the local maximal amplitudes, agree well with the predictions of Eq. (22) for the local most-probable route.
At the same time, the curves of classical limit of Eq. (33) are located in the corresponding  inner rings of PMDs,
while the curves of partial-decoupling approximation of Eq. (34) are located in the corresponding outer rings.
By comparing Fig. 5(a) with Fig. 5(b), we can also see that with the increase of the laser wavelength, 
the local most-probable route of Eq. (22) is closer to the classical limit of Eq. (33).
These results are consistent with those in OTC fields, indicating that our conclusion is also applicable to cases of
2D tunneling ionization in circular laser fields.

\begin{figure}[t]
\begin{center}
\rotatebox{0}{\resizebox *{8.5cm}{8.5cm} {\includegraphics {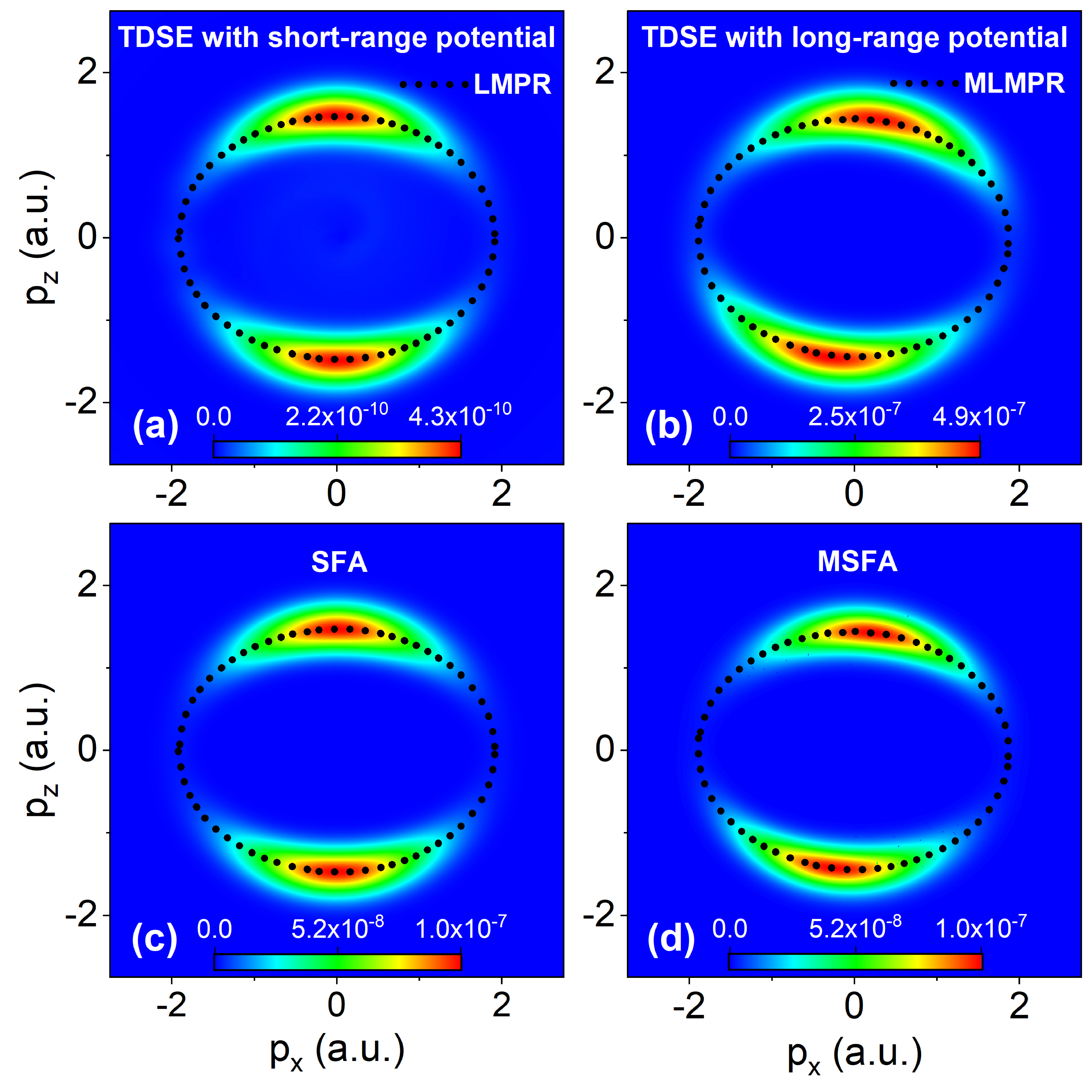}}}
\end{center}
\caption
{(Color online)  PMDs of model He atom in  elliptical laser fields with $\epsilon=0.8$, obtained with TDSE of short-range Coulomb potential (a) and long one (b),
SFA (c) and modified SFA (MSFA) considering  effects of long-range Coulomb potential (d).
The peak intensity of the elliptical laser field is $I=5\times10^{14}\text{W/cm}^{2}$ and the wavelength is $\lambda=800$ nm.
The predictions of Eq. (22) for the local most-probable route (LMPR) (dotted) are plotted in (a) and (c), and the predictions of the Coulomb-modified LMPR (MLMPR) (dotted) are plotted in (b) and (d).
} \label{Fig. 6}
\end{figure}

\section{Effects of Coulomb potential}
In the above SFA-based discussions, we have neglected the Coulomb effect.
The Coulomb effect indeed can change PMDs of SFA predictions \cite{Goreslavski,TMYan2012} and accordingly influence the local most-probable tunneling routes.
For cases discussed above, this influence is not noticeable, with the agreement between TDSE and SFA predictions.
This influence can be non-negligible for 2D laser fields with other forms such as elliptical ones \cite{Yan2020} and OTC fields with the relative phase $\phi={\pi}/{2}$ \cite{Busulad2020}.

Here, we discuss the effects of Coulomb potential on the local most-probable route. It is well known that in elliptical laser fields,
the Coulomb effect will induce the rotation of the PMD. In Fig. 6, we present relevant comparisons for model He atoms with long-range and short-range potentials.
The long-rang potential has the form as introduced below Eq. (1). The short one has the form as introduced in  \cite{chen2019}.
One can observe that the PMD of TDSE simulations with the short-range potential in Fig. 6(a) is symmetric with respect to the axis of $p_x=0$, which is
similar to the SFA result in Fig. 6(c). By contrast, the PMD of TDSE simulations with the long-range potential in Fig. 6(b) shows a remarkable rotation.
With using a modified SFA (MSFA) model that considers the effect of the long-range potential \cite{chen2019}, this rotation is reproduced in Fig. 6(d).
With using Eq. (22), we can also obtain
the local most-probable routes for the elliptical case, which agree well with the TDSE results  of short-range potential, as seen in Fig. 6(a).
With propagating the electron trajectories associated with these local most-probable routes of Eq. (22) using MSFA,
the Coulomb-modified local most-probable route also gives a good description of this rotation, as seen in Figs. 6(b) and 6(d).

\begin{figure}[t]
\begin{center}
\rotatebox{0}{\resizebox *{8.5cm}{8.5cm} {\includegraphics {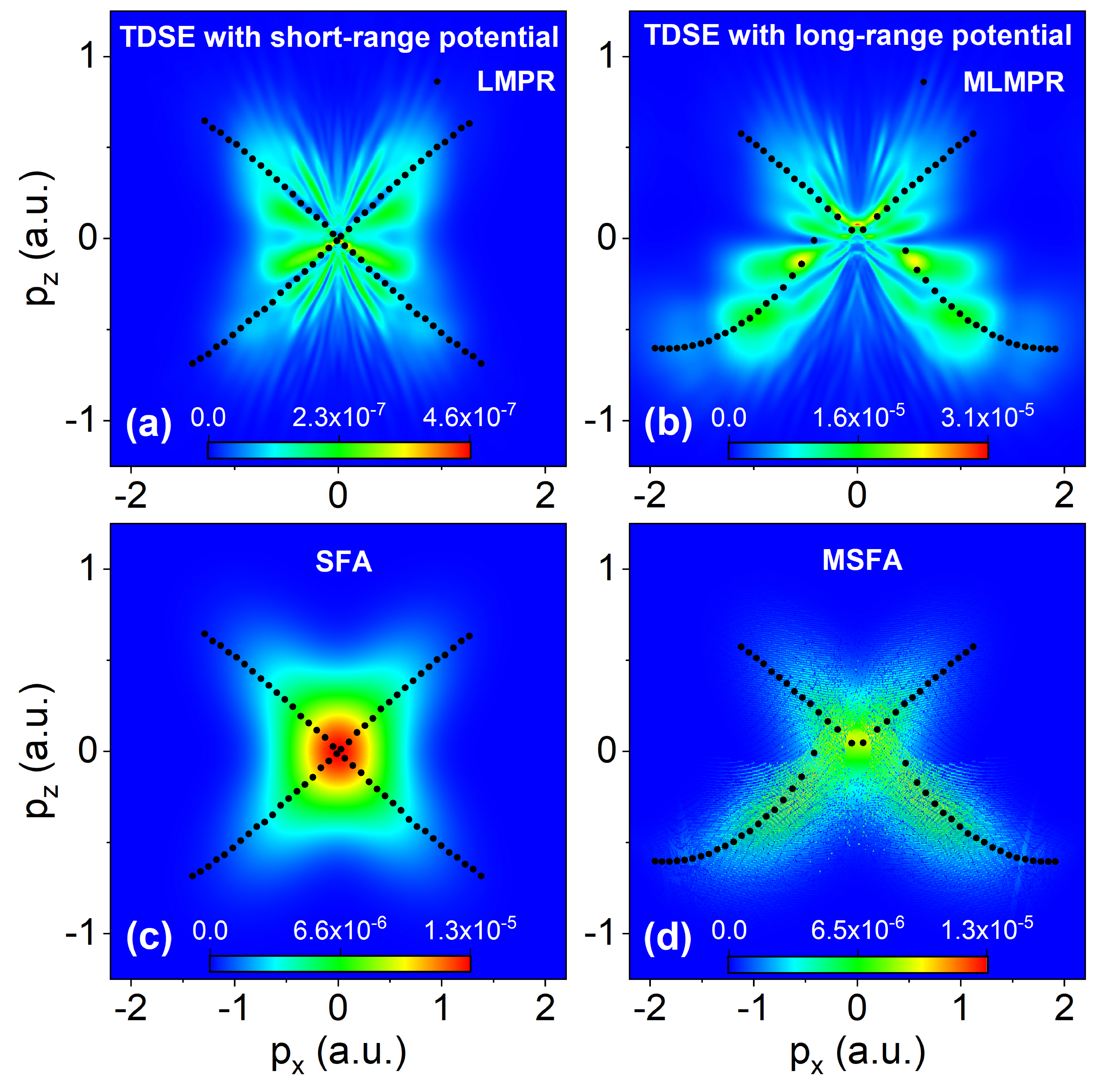}}}
\end{center}
\caption
{(Color online)
Same as Fig. 6, but for OTC laser fields of $\phi={\pi}/{2}$, with $I_1=5\times10^{14}\text{W/cm}^2$, $\lambda_1=800$ nm, and $\varepsilon=0.5$.
} \label{Fig. 7}
\end{figure}

Further comparisons for  OTC laser fields with $\phi=\pi/2$ are presented in Fig. 7. For the present cases, one can observe that
without considering the Coulomb effect, PMDs of TDSE simulations with short-range potential and SFA show a  basic symmetric (TDSE) or symmetric (SFA) structure with respect
to the axis of $p_z=0$, and Eq. (22) gives an applicable description for the local maximal amplitudes of these PMDs, as seen in Figs. 7(a) and 7(c).
However, the PMD of TDSE simulations with long-range potential shows a  structure with striking asymmetry with respect to $p_z=0$ and this striking asymmetry
is reproduced by MSFA, as shown in Figs. 7(b) and 7(d). The origin of this striking asymmetry has been attributed to the Coulomb induced ionization time lag,
which is discussed in detail in \cite{chen2019}.
In this case, one can see from the right column of Fig. 7,  the curves of the Coulomb-modified local most-probable routes related to Eq. (22) and MSFA,
also agree with the bright parts of these asymmetric PMDs.

We mention that there exits the massive disagreement between TDSE and SFA for the short-range potential in Fig. 7. The reason can be as follows.
In the SFA simulations, we only consider the contribution of direct ionization related to saddle points of Eq. (7), where the rescattering of the electron by the parent ion is neglected.
This rescattering process exits in the TDSE simulations even for the case of short-range potential.
The interference of direct electron and rescattering electron will generate interference fringes in PMD, as shown in Figs. 7(a) and 7(b).
This interference effect is absent in our SFA simulations.
By comparison, this interference effect is partly considered in our MSFA simulations where the Coulomb potential is included and can induce the rescattering of the electron.
Moreover, in this case, the local most-probable route does not match the maxima of the PMD very well.
The reason can be that for the case of the relative OTC phase $\phi=90^o$, the rescattering process is more likely to happen due to the regulation of the second-harmonic field on the electron's motion.
As a result, the local most-probable route, which is obtained with the analysis of saddle points of Eq. (7) without considering the rescattering effect,
deviates somewhat from the local maximum of the PMD. In addition, in this case, the PMD has large amplitudes around small momenta near zero.
For larger momenta, the amplitudes are similar. Consequently, the local most-probable route related to the local maximal amplitude in PMD is also somewhat more difficult to identify.
We therefore anticipate that the condition of Eq. (22) is more applicable for cases where the rescattering effect is weak.

\begin{figure}[t]
\begin{center}
\rotatebox{0}{\resizebox *{8.7cm}{6.8cm} {\includegraphics {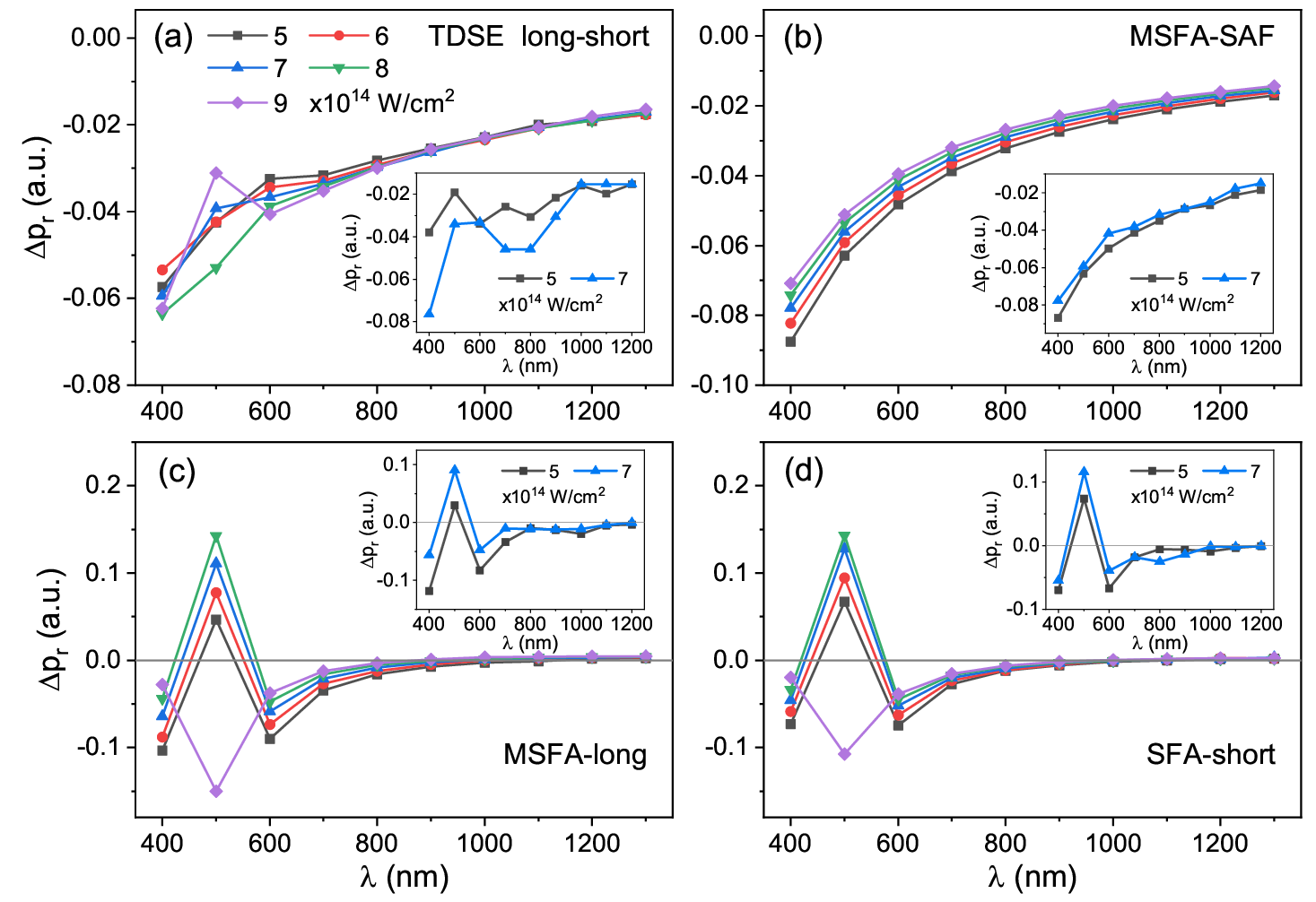}}}
\end{center}
\caption
{(Color online)
Comparisons of the average most-probable radius $p_r$ in ring-shaped  PMDs of circular laser fields
at different laser intensities and wavelengthes, obtained with TDSE of long-range versus short-range Coulomb potentials, SFA and MSFA.
(a):  Results of TDSE with long-range potential minus short ones. (b): Results of MSFA minus SFA ones. (c): Results of MSFA minus those of TDSE with long-range potential.
(d): Results of SFA minus those of TDSE with short-range potential. The TDSE average radius is calculated with dividing the PMD into small sectors with 1-degree central angle and
finding the local most-probable radius related to the local maximal amplitude in each sector, then averaging the obtained radius of all sectors with weight. The model one is calculated
with averaging the radius of the local most-probable route of Eq. (22)  with weight.
The insets in each panel show the corresponding results of the most-probable radius
(which is related to the maximal amplitude in PMD).} \label{Fig. 8}
\end{figure}

\section{Further considerations of Coulomb effect}
In strong-field  ultrafast experiments with using 2D laser fields to measure and control the motion of electrons within atoms and molecules,
one hopes to retrieve the dynamics or structure information of the target from experimental observables such as PMDs.
To do so, applicable theoretical models are needed to build a bridge between experimental observables and the desired information.
For cases of elliptically-polarized laser fields, the most bright part of PMD, associated with the most-probable route and with remarkably larger amplitudes
than other parts, is well defined in the distribution.
One therefore can use the most bright part of PMD as the preferred observable, instead of the total momentum distribution, to deduce the desired information quantitatively.
For example, in attosecond clock experiments, the most bright part in PMDs has been used
to obtain the offset angle. With comparing this angle obtained in experimental measurements
with that obtained with semiclassical simulations
(where the tunneling event is considered to occur instantaneously), one is able to access the electron motion under the laser-Coulomb-formed barrier
and explore the issue of the tunneling time (i.e., if a real time is needed when  the electron  tunnels through the barrier).

For other cases of 2D laser fields such as OTC laser fields with different delay between these two colors or circularly-polarized laser fields, the situation is different.
The most bright part in PMDs is usually not well defined  for cases of OTC laser fields (see Fig. 1 and Fig. 7), and it shows a ring-shaped distribution  for circular laser fields  (see Fig. 5).
For the cases, the distribution associated with the local most-probable route explored in this paper can be the preferred observable in quantitative deduction, instead of the most-probable one.
To illuminate this point, next, we give a concrete case for circular laser fields which have shown important applications in ultrafast control of 
electron motion \cite{Silva}. The circular laser field  with high time and space symmetry also allows us to explore some basic aspects of strong-field physics such as wavelength scaling of
Coulomb effect after the electron tunnels out of the barrier and wavelength dependence of electron tunneling dynamics under the barrier.

In Fig. 8, we compare the average most-probable radius (AMPR) of momentum $p_r=\sqrt{p_x^2+p_z^2}$ in ring-shaped PMDs
of circular laser fields for diverse laser parameters, obtained with different methods.
The AMPR is evaluated with averaging the radius associated with the local most-probable route (which is corresponding to the local maximal amplitude in ring-shaped PMD) over different azimuths. 
We consider TDSE simulations with both long-range and short-range Coulomb potentials as well as SFA and MSFA simulations.
First of all, in Fig. 8(a), we show the difference of AMPR between TDSE of long-range versus short-range potentials.
This difference is negative, implying that  the long-range potential decreases the AMPR on the whole, in comparison with the short-range one.
Results at different laser intensities differ remarkably for short laser wavelengthes and become regular and comparable for wavelengthes longer than 600 nm.
This difference of AMPR decreases with increasing the laser wavelength, and for long wavelengthes such as 1200 nm, the results of different laser intensities coincide with each other.
The behaviors of the TDSE curves for wavelengthes longer than 600 nm in Fig. 8(a) are basically reproduced by the SFA models in Fig. 8(b).
Considering that the model calculations include the contributions of only the ground state to ionization,
the irregular behaviors of the TDSE curves at wavelengthes shorter than 600 nm can arise from the effect of the excited state.
In the following discussions, we focus on cases of laser wavelengthes longer than 600 nm.

The results in Figs. 8(a) and 8(b) provide deep insight into the effect of the long-range Coulomb potential on the tunneling-out electron.
1) In comparison with the laser intensity, the Coulomb effect is more sensitive to the laser wavelength.
2) The Coulomb effect decreases remarkably with increasing the laser wavelength.
One of the possible reasons for these phenomena in Fig. 8(a) of TDSE simulations  can be that for longer laser wavelengthes,
the position of the tunnel exit is farther away from the nucleus and
accordingly the influence of the Coulomb potential on the motion of the electron after it exits the barrier is smaller.
The potential mechanism of these phenomena can be further explored with analyzing the model results in Fig. 8(b)
associated with the local most-probable route defined with Eq. (22), which deserves a detailed study in the future.

Other insights into the dynamics of the tunneling electron under the  barrier are revealed in the second row of Fig. 8,
where we show the difference of AMPR between MSFA and TDSE of long-range potential as well as that between SFA and TDSE of short-range potential for different laser parameters.
For wavelengthes longer than 600 nm, the curves in Fig. 8(c) also show the regular behavior. This difference is negative and decreases with increasing the laser wavelength.
It becomes near to zero for wavelengthes longer than 900 nm. In addition, for a certain laser wavelength, this difference decreases with the increase of the laser intensity.
The phenomena are reproduced in Fig. 8(d) where predictions of SFA and TDSE of short-range potential for AMPR are compared,
implying that the long-range Coulomb potential plays a small role in the phenomena in Fig. 8(c).
The SFA with the saddle-point theory describes the tunneling process under the barrier in term of the imaginary time.
The difference of AMPR revealed in Figs. 8(c) and 8(d) implies that the tunneling process at relatively short laser wavelengthes
(with the Keldysh parameter near or larger than 1 for the present laser intensities) is more complex beyond the description of SFA.
One of the possible mechanisms for the short-wavelength cases is excitation tunneling \cite{Chen2011,Sereb}
where the ground-state electron is first pumped into an excited state then it ionizes through tunneling from the excited state.
In this meaning, the difference of AMPR revealed in Figs. 8(c) and 8(d) characterizes the complex tunneling dynamics associated with multiphoton excitation.

The rich information revealed in Fig. 8, however, can not be fully accessed with the most-probable radius in PMDs, as shown in the insets in Fig. 8.
In the insets, for clarity, we show results only for two laser intensities of $I=5\times10^{14}\text{W/cm}^2$ and $I=7\times10^{14}\text{W/cm}^2$.
The results of the most-probable radius show some oscillation for diverse laser wavelengthes,
precluding an accurate identification of the wavelength-dependent law of relevant phenomena.
We mention that for the cases of circular laser fields discussed above, due to the envelop of the laser pulse or the numerical instability,
the local most-probable radiuses evaluated differ somewhat from each other. It is the reason
that the AMPR obtained with weight averaging is preferred in measurements instead of the most-probable radius here.

\section{Extended discussions}

It should be stressed that in this paper, we focus our discussions on atoms. For molecules with multi-center structure, some complex effects emerge in strong-field ionization,
such as quantum interference between the molecular centers during tunneling \cite{Chen2010,Chencj2012,Kunitski2019},
permanent dipole induced large Stark effect \cite{Bandrauk2,Madsen2010,Wang2019}, laser induced nuclear quick stretching  \cite{lein2,Baker,Li2016,Lan2017,Li2019},
and  permanent dipole induced direct vibration excitation \cite{Wustelt,Yue,wang2020}, etc..
For the purpose of achieving more precise control and measurement of the electron motion within the molecule with 2D laser fields,
studies on influences of molecular structure and structure-related effects on properties of 2D tunneling are also expected.

\section{Conclusion}
In conclusion, we have studied  tunneling ionization of model atoms in
intense 2D laser fields such as orthogonally-polarized two-color ones and circularly-polarized ones.
In a semiclassical picture relating to SFA-based electron trajectories, the electron which tunnels out of the laser-Coulomb-formed barrier, usually has
a non-vanishing imaginary part of the exit position.
We have shown that
tunneling in 2D laser fields is preferred as this imaginary part is small.
A condition for the local most-probable tunneling route which corresponds to the regions of large amplitude in PMD is obtained.
It characterizes the properties of relevant routes and can be used in analytical treatments.

We have also compared the local most-probable route with the predictions of the classical limit,
and the partial-decoupling approximation where it is assumed that the main component of the 2D laser field with large amplitude
dominates in tunneling. For the same longitude momentum $p_x$, when the classical limit underestimates the
corresponding transverse momentum $p_z$, the partial-decoupling approximation overestimates that,
with revealing the important modulation of the minor component of the 2D field in tunneling.
As the Keldysh parameter decreases,  this local most-probable route begins to approach and finally agrees with the classical one.
Our approaches for identifying the local most-probable route can also be used for cases
where the Coulomb effect is marked such as ATI in elliptical laser fields and OTC fields with larger time delay.

In ultrafast experiments with using 2D laser fields to measure and control the motion of electrons within atoms and molecules,
one quantitatively deduces the structure or dynamics information of the target from experimental observables with the help of theory model.
The region of the maximal amplitude in PMD associated with  the most-probable route has been often used as a preferred observable in experiments.
The regions of large amplitude in PMD related to this local most-probable route identified in the paper, however, provide another choice, especially for
the cases where the region of the maximal amplitude in PMD is not well defined or
weight averaging is needed to 
distill information from observable data of PMD with high accuracy.
For example, using the regions of large amplitude in PMD of circular laser fields as the observable, one can explore
some quantitative characteristics of strong-field tunneling ionization in different timing,
such as wavelength scaling of Coulomb effect after the tunneling electron exits the barrier and wavelength dependence of dynamics of the tunneling electron under the barrier.
This local most-probable route identified here  provides a theoretical tool for further understanding and retrieving information from the characteristics.
Our work provides a perspective for studying the complex dynamics of tunneling in strong 2D laser fields.

\section*{Acknowledgements}
This work was supported by the National Natural Science Foundation of China (Grants  No. 11904072 and No. 91750111),
the National Key Research and Development Program of China (Grant No. 2018YFB0504400),
Scientific research program of Education Department of Shaanxi Provience, China (18JK0098), and Scientific Research Foundation of SUST, China (2017BJ-30).

\begin{widetext}
\section{Appendix}
\subsection{}
Inserting the solution of Eq. (9) into the semiclassical action $S(\mathbf{p},t')$, one can get the imaginary part of the semiclassical action
\begin{eqnarray}
\begin{aligned}
S^\text{Im}(\mathbf{p},t'_s)=&\frac{U_{p1}}{2\omega_1}\text{cos}(2\omega_1 t'_{sr})\text{sinh}(2\omega_1 t'_{si})
+\frac{U_{p2}}{2\omega_2}\text{cos}(2\omega_2 t'_{sr}+2\phi)\text{sinh}(2\omega_2 t'_{si})
+p_x\frac{A_1}{\omega_1}\text{cos}(\omega_1 t'_{sr})\text{sinh}(\omega_1 t'_{si})\\
&+p_z\frac{A_2}{\omega_2}\text{cos}(\omega_2 t'_{sr}+2\phi)\text{sinh}(\omega_2 t'_{si})+(\frac{\mathbf{p}^2}{2}+I_p+U_{p1}+U_{p2})t'_{si}.
\end{aligned}
\end{eqnarray}
The corresponding term $\text{e}^{-S^\text{Im}(\mathbf{p},t'_s)}$ is related to the amplitude of the photoelectron with the drift momentum $\mathbf{p}$. The real part of the action is
\begin{eqnarray}
\begin{aligned}
S^\text{Re}(\mathbf{p},t'_s)=&\frac{U_{p1}}{2\omega_1}\text{sin}(2\omega_1 t'_{sr})\text{cosh}(2\omega_1 t'_{si})
+\frac{U_{p2}}{2\omega_2}\text{sin}(2\omega_2 t'_{sr}+2\phi)\text{cosh}(2\omega_2 t'_{si})
+p_x\frac{A_1}{\omega_1}\text{sin}(\omega_1 t'_{sr})\text{cosh}(\omega_1 t'_{si})\\
&+p_z\frac{A_2}{\omega_2}\text{sin}(\omega_2 t'_{sr}+2\phi)\text{cosh}(\omega_2 t'_{si})
+(\frac{\mathbf{p}^2}{2}+I_p+U_{p1}+U_{p2})t'_{sr}+S_{t_F}.
\end{aligned}
\end{eqnarray}
The corresponding term $\text{e}^{iS^\text{Re}(\mathbf{p},t'_s)}$ is related to the phase of the photoelectron.
Note that, the value of the term
$S_{t_F}=-\frac{U_{p1}}{2\omega_1}\text{sin}(2\omega_1 t_F)-\frac{U_{p2}}{2\omega_2}\text{sin}(2\omega_2 t_F+2\phi)
-p_x\frac{A_1}{\omega_1}\text{sin}(\omega_1 t_F)-p_z\frac{A_2}{\omega_2}\text{sin}(\omega_2 t_F+2\phi)
-(\frac{\mathbf{p}^2}{2}+I_p+U_{p1}+U_{p2})t_F$
depends on the final time $t_F$ and is constant for every trajectory.
Therefore, this term can be considered as a phase shift related to the final time $t_F$,
and plays no role in the interference between the trajectories.

It is well known that, for the case of $\phi=0$, there is a pair of saddle points in one laser cycle for the same drift momentum $\mathbf{p}$
and the interference between these two trajectories contributes to the interference fringes in PMD.
The relation between these two saddle points in the n-$th$ cycle is
\begin{eqnarray}
\left\{
\begin{aligned}
t'_{s2r}=&nT-t'_{s1r},\\
t'_{s2i}=&t'_{s1i}.
\end{aligned}
\right.
\end{eqnarray}
Therefore, we have
\begin{eqnarray}
\left\{
\begin{aligned}
\text{cos}(\omega_1 t'_{s2r})&=\text{cos}(\omega_1 t'_{s1r}),\\
\text{sin}(\omega_1 t'_{s2r})&=-\text{sin}(\omega_1 t'_{s1r}),\\
\text{cosh}(\omega_1 t'_{s2i})&=\text{cosh}(\omega_1 t'_{s1i}),\\
\text{sinh}(\omega_1 t'_{s2i})&=\text{sinh}(\omega_1 t'_{s1i}).\\
\end{aligned}
\right.
\end{eqnarray}
Then we arrive at the relation $S^\text{Im}(\mathbf{p},t'_{s2})=S^\text{Im}(\mathbf{p},t'_{s1})$,
which shows that the amplitudes of the pair of trajectories are the same.
At the same time, the phase difference $\Delta \theta_{\mathbf{p}}$ of these two trajectories is
\begin{eqnarray}
\begin{aligned}
\Delta \theta_{\mathbf{p}}=&S^\text{Re}(\mathbf{p},t'_{s1})-S^\text{Re}(\mathbf{p},t'_{s2})\\
=&\frac{U_{p1}}{\omega_1}\text{sin}(2\omega_1 t'_{s1r})\text{cosh}(2\omega_1 t'_{s1i})
+\frac{U_{p2}}{\omega_2}\text{sin}(2\omega_2 t'_{s1r})\text{cosh}(2\omega_2 t'_{s1i})
+\frac{2A_1p_x}{\omega_1}\text{sin}(\omega_1 t'_{s1r})\text{cosh}(\omega_1 t'_{s1i})\\
&+\frac{2A_2p_z}{\omega_2}\text{sin}(\omega_2 t'_{s1r})\text{cosh}(\omega_2 t'_{s1i})
+(\frac{\mathbf{p}^2}{2}+I_p+U_{p1}+U_{p2})[2t'_{s1r}-2nT].
\end{aligned}
\end{eqnarray}
Note that the above expression is applicable for $t'_{s1r}\in (0.25T+nT,0.5T+nT)$. If $t'_{s1r}\in (nT,0.25T+nT]$,
one should move the value of $t'_{s2r}$ forward one cycle. That is $t'_{s2r}=(n-1)T-t'_{s1r}$. In this case, the relation $S^\text{Im}(\mathbf{p}, t'_{s2})=S^\text{Im}(\mathbf{p}, t'_{s1})$  still works.
However, the phase difference $\Delta \theta_{\mathbf{p}}$ becomes
\begin{eqnarray}
\begin{aligned}
\Delta \theta_{\mathbf{p}}=&S^\text{Re}(\mathbf{p},t'_{s1})-S^\text{Re}(\mathbf{p},t'_{s2})\\
=&\frac{U_{p1}}{\omega_1}\text{sin}(2\omega_1 t'_{s1r})\text{cosh}(2\omega_1 t'_{s1i})
+\frac{U_{p2}}{\omega_2}\text{sin}(2\omega_2 t'_{s1r})\text{cosh}(2\omega_2 t'_{s1i})
+\frac{2A_1p_x}{\omega_1}\text{sin}(\omega_1 t'_{s1r})\text{cosh}(\omega_1 t'_{s1i})\\
&+\frac{2A_2p_z}{\omega_2}\text{sin}(\omega_2 t'_{s1r})\text{cosh}(\omega_2 t'_{s1i})
+(\frac{\mathbf{p}^2}{2}+I_p+U_{p1}+U_{p2})[2t'_{s1r}-(2n-1)T].
\end{aligned}
\end{eqnarray}
When $\Delta \theta_{\mathbf{p}}=(2q+1)\pi$ ($\Delta \theta_{\mathbf{p}}=2q\pi$)
one can observe the destructive (constructive) interference in PMD.

\subsection{}
For the case of $\phi=0$, inserting the imaginary parts of Eq. (13)
\begin{equation}
\left\{
\begin{aligned}
A_x^\text{Im}(t'_s)=&-A_1\text{sin}(\omega_1 t'_{sr})\text{sinh}(\omega_1 t'_{si}),\\
A_z^\text{Im}(t'_s)=&-A_2\text{sin}(\omega_2 t'_{sr}+\phi)\text{sinh}(\omega_2 t'_{si})
\end{aligned}
\right.
\end{equation}
into the second term of Eq. (14)
\begin{eqnarray}
[A_x^\text{Im}(t'_s)]^2+[A_z^\text{Im}(t'_s)]^2=2I_p,
\end{eqnarray}
one can get the expression
\begin{eqnarray}
A_1^2\text{sin}^2(\omega_1 t'_{sr})\text{sinh}^2(\omega_1 t'_{si})+A_2^2\text{sin}^2(\omega_2 t'_{sr})\text{sinh}^2(\omega_2 t'_{si})=2I_p.
\end{eqnarray}
With the double-angle formulas of cos and cosh functions, Eq. (44) can be written as
\begin{equation}
\begin{aligned}
&A_1^2\text{sin}^2(\omega_1 t'_{sr})\text{sinh}^2(\omega_1 t'_{si})+A_2^2\text{sin}^2(\omega_2 t'_{sr})\text{sinh}^2(\omega_2 t'_{si})\\
=&\text{sin}^2(\omega_1 t'_{sr})\text{sinh}^2(\omega_1 t'_{si})[A_1^2+16A_2^2\text{cos}^2(\omega_1 t'_{sr})\text{cosh}^2(\omega_1 t'_{si})]\\
=&[1-\text{cos}^2(\omega_1 t'_{sr})][\text{cosh}^2(\omega_1 t'_{si})-1][A_1^2+16A_2^2\text{cos}^2(\omega_1 t'_{sr})\text{cosh}^2(\omega_1 t'_{si})]\\
=&[\text{cos}^2(\omega_1 t'_{sr})+\text{cosh}^2(\omega_1 t'_{si})-\text{cos}^2(\omega_1 t'_{sr})\text{cosh}^2(\omega_1 t'_{si})-1][A_1^2+16A_2^2\text{cos}^2(\omega_1 t'_{sr})\text{cosh}^2(\omega_1 t'_{si})]\\
=&[\text{cos}^2(\omega_1 t'_{sr})+\text{cosh}^2(\omega_1 t'_{si})-\frac{p_x^2}{A_1^2}-1](A_1^2+16A_2^2\frac{p_x^2}{A_1^2})\\
=&2I_p.
\end{aligned}
\end{equation}
According to
\begin{equation}
\begin{aligned}
&\text{cos}(\omega_2 t'_{sr})\text{cosh}(\omega_2 t'_{si})\\
=&[2\text{cos}^2(\omega_1 t'_{sr})-1][2\text{cosh}^2(\omega_1 t'_{si})-1]\\
=&4\text{cos}^2(\omega_1 t'_{sr})\text{cosh}^2(\omega_1 t'_{si})-2[\text{cos}^2(\omega_1 t'_{sr})+\text{cosh}^2(\omega_1 t'_{si})]+1\\
=&4\frac{p_x^2}{A_1^2}-2[\text{cos}^2(\omega_1 t'_{sr})+\text{cosh}^2(\omega_1 t'_{si})]+1\\
=&-\frac{p_z}{A_2},
\end{aligned}
\end{equation}
we get
\begin{equation}
\begin{aligned}
\text{cos}^2(\omega_1 t'_{sr})+\text{cosh}^2(\omega_1 t'_{si})=\frac{1}{2}(4\frac{p_x^2}{A_1^2}+\frac{p_z}{A_2}+1).
\end{aligned}
\end{equation}
Then inserting Eq. (47) into (45), we have
\begin{equation}
\begin{aligned}
{[\frac{1}{2}(4\frac{p_x^2}{A_1^2}+\frac{p_z}{A_2}+1)-\frac{p_x^2}{A_1^2}-1]}(A_1^2+16A_2^2\frac{p_x^2}{A_1^2})=2I_p.
\end{aligned}
\end{equation}
The above expression gives the relation between $p_x$ and $p_z$ in the partial-decoupling approximation, and can be further simplified as
\begin{eqnarray}
\begin{aligned}
p_z=A_2(\frac{I_p}{U_{p1}+4\frac{U_{p2}}{U_{p1}}p_x^2}-\frac{p_x^2}{2U_{p1}}+1).
\end{aligned}
\end{eqnarray}
In the classical limit, according to Eq. (16)
\begin{eqnarray}
\left\{
\begin{aligned}
p_x=&-A_1\text{cos}(\omega_1 t')\\
p_z=&-A_2\text{cos}(\omega_2 t')=-A_2[2\text{cos}^2(\omega_1 t')-1],
\end{aligned}
\right.
\end{eqnarray}
we can get the following relation between $p_x$ and $p_z$
\begin{eqnarray}
\begin{aligned}
p_z=A_2(-\frac{p_x^2}{2U_{p1}}+1).
\end{aligned}
\end{eqnarray}

\end{widetext}

\end{document}